\documentstyle{article}

\begin{document}
\baselineskip = 20pt
\input epsf
\ifx\epsfbox\UnDeFiNeD\message{(NO epsf.tex, FIGURES WILL BE IGNORED)}
\def\figin#1{\vskip2in}
\else\message{(FIGURES WILL BE INCLUDED)}\def\figin#1{#1}\fi

\def\ifig#1#2#3{\xdef#1{fig.~\the\figno}
\goodbreak\figin{\centerline{#3}}%
\smallskip\centerline{\vbox{\baselineskip12pt
\advance\hsize by -1truein\noindent{\bf
Fig.~\the\figno:} #2}}
\bigskip\global\advance\figno by1}

\def\figures{\centerline{{\bf Figure
Captions}}\medskip\parindent=40pt%
\def\fig##1##2{\medskip\item{FIG.~##1.  }##2}}
\newwrite\ffile\global\newcount\figno \global\figno=1
\def\fig{fig.~\the\figno\nfig}
\def\nfig#1{\xdef#1{fig.~\the\figno}%
\writedef{#1\leftbracket fig.\noexpand~\the\figno}%
\ifnum\figno=1\immediate\openout\ffile=figs.tmp\fi\chardef\wfile=
\ffile%
\immediate\write\ffile{\noexpand\medskip\noexpand\item{Fig.\
\the\figno. }
\reflabeL{#1\hskip.55in}\pctsign}\global\advance\figno by1\findarg}

\parindent 25pt
\overfullrule=0pt
\tolerance=10000
\def\Re{\rm Re}
\def\Im{\rm Im}
\def\titlestyle#1{\par\begingroup \interlinepenalty=9999
     \fourteenpoint
   \noindent #1\par\endgroup }
\def\tr{{\rm tr}}
\def\Tr{{\rm Tr}}
\def\half{{\textstyle {1 \over 2}}}
\def\quart{{\textstyle {1 \over 4}}}
\def\calt{{\cal T}}
\def\ie{{\it i.e.}}
\def\np{Nucl. Phys.}
\def\pl{Phys. Lett.}
\def\pr{Phys. Rev.}
\def\prl{Phys. Rev. Lett.}
\def\cmp{Comm. Math. Phys.}
\def\quart{{\textstyle {1 \over 4}}}
\def\RR{${\rm R}\otimes{\rm R}~$}
\def\NSNS{${\rm NS}\otimes{\rm NS}~$}
\def\RNS{${\rm R}\otimes{\rm NS}~$}
\def\calf{${\cal F}$}

\baselineskip=14pt
\pagestyle{empty}
{\hfill DAMTP/96-104}
\vskip 0.1cm
{\hfill hep-th/9701093}
\vskip 0.4cm
\centerline{EFFECTS OF D-INSTANTONS.}
\vskip 1cm
 \centerline{ Michael B.  Green\footnote{M.B.Green@damtp.cam.ac.uk}
 and Michael Gutperle\footnote{M.Gutperle@damtp.cam.ac.uk}}
\vskip 0.3cm
\centerline{DAMTP, Silver Street,}
\centerline{ Cambridge CB3 9EW, UK.}
\vskip 1.4cm
\centerline{ABSTRACT}
\vskip 0.3cm

Scattering of  fundamental  states of  type IIB supergravity and
superstring
theory  is discussed at low orders in
perturbation theory  in the background of a D-instanton.  The
integration over
fermionic zero modes   in both the low energy  supergravity and in
the string
theory  leads to explicit
nonperturbative terms in  the effective action.  These include  a single
instanton correction to the known tree-level and one-loop  $R^4$
interactions.
The `spectrum' of multiply-charged  D-instantons is deduced by
T-duality in
nine dimensions from multiply-wound world-lines of marginally-bound
D-particles.
This, and other clues,  lead to a conjectured   SL(2,Z) completion
of the $R^4$
terms which suggests  that they are  not renormalized by perturbative
corrections in the zero-instanton sector beyond one loop.
The string  theory unit-charged D-instanton gives rise to
point-like effects in
 fixed-angle scattering, raising unresolved issues concerning
distance scales
 in superstring theory.

\vfill\eject
\pagestyle{plain}
\setcounter{page}{1}

\section{Introduction}

The  dualities  that relate   superstring perturbation
expansions  (and M-theory)  involve the
interchange of fundamental string excitations and solitons.  In
compactified theories  these symmetries relate the different states
obtained by
wrapping both the strings and their solitons  around compact
dimensions, but
already in ten
dimensions there are interesting inter-relationships between theories.
Notably, the type IIB theory is self-dual in the sense that it
transforms into
itself under   $SL(2,Z)$  transformations which act  on
 the various species of solitons -- the $SL(2,Z)$ multiplet of
strings, the
self-dual three-brane, the $SL(2,Z)$ multiplet
of fivebranes and the seven-brane.   However, the type IIB theory also
possesses an instanton solution which couples to the Ramond--Ramond (\RR)
charge associated with the pseudoscalar, $C^{(0)}$.\footnote{We are
using the
convention that the \RR $q$-form potentials are denoted $C^{(q)}$.}  This
leads to little-studied non-perturbative effects in the
ten-dimensional theory
-- effects that are intimately related to the $SL(2,Z)$ symmetry of
the theory.

After compactification there are other instantons, namely,   BPS   \lq
$(p+1)$-instantons'   that arise from  wrapping the $(p+1)$-dimensional
euclidean world-volumes  of  $p$-branes around compact dimensions  in a
supersymmetric manner  \cite{stromingerbecker, ooguri,
harveymoore}.   All such
instantons, which are related to each other by T-duality,  have infinite
actions in the limit of  flat ten-dimensional space-time  apart  from the
D-instanton ($p=-1$), which has
a zero-dimensional world-volume.   Such euclidean configurations
 may be treated in a unified manner, for example, using the formalism of
\cite{greengutperlec}.

  Although some  properties of these $(p+1)$-instantons have been
studied  and
their duality properties have  been explored, relatively
little is known about correlation functions in the instanton
backgrounds and
even less  about the  effect of the D-instanton on  ten-dimensional type
IIB theory.   It would, for example,  be  interesting to study the
$SL(2,Z)$
extension of the perturbative effective action, which might reveal
connections
with more fundamental underlying principles.   It is  particularly
fascinating
that D-instantons  appear to give  rise to point-like  behaviour of
short-distance
correlation functions of fundamental strings.  Indeed the
point-like nature of
Dirichlet boundary conditions has for some time been linked to
power behaviour
of  fixed-angle scattering in the bosonic string theory
\cite{greenpointlike,greendual,gutperlea}.   This is in marked
contrast to the
exponential
decrease of fixed-angle cross sections that is characteristic of
conventional
fundamental string processes \cite{veneziano,amati,grossmende}.

This paper is concerned with  effects of   D-instantons in IIB
supergravity and  superstring theory in ten uncompactified dimensions.
Much of the analysis also applies to any of the  $(p+1)$-instantons
that arise
in compactifications to lower
dimensions.  We will begin by reviewing the  D-instanton solution of
IIB supergravity in section 2. An expanded discussion of the instanton
action will be given in the appendix. Just as in the case of   
D-branes with
$p>-1$ the  classical supergravity solution \cite{greenc} is
singular at the
origin,
but the string theory description is expected to be well-defined
at short
distances \cite{wittenboundstates,pouliot,ferretti,douglas}.  The
fermionic
zero modes of the IIB supergravity field theory  in this
background are  obtained in section 3  by  applying the generators of the
sixteen broken supersymmetries to the instanton
field configuration.   The  counterpart of these   modes in the IIB
superstring  will be obtained  in section 4 where the fermionic
zero modes
are given by attaching fermionic open string vertex operators to the
boundary of world sheets.
Integration over the sixteen fermionic moduli   induces new interactions,
just such as it does in conventional field theory \cite{thooft}.

In section 5 we will discuss interactions that are induced at
leading order in
an instanton background   in both the supergravity and the string
descriptions.
  These include a    $\lambda^{16}$  term  and a $R^4$  term (where
$R$ is the
Riemann curvature).  The latter  is a one-instanton correction to
the tree
level \cite{grisaru,wittengross} and one loop  
\cite{greenschwarziii}  $R^4$
terms  with
the same tensor structure. These particular terms are non-stringy
in origin
since they can be deduced from IIB supergravity alone.

The explicit calculations of this paper are limited to  a single
unit-charged
D-instanton background since little is known about multiply-charged
D-instantons beyond the classical solutions of IIB supergravity.
However,
T-duality provides more detailed information.  Thus,   T-duality
in nine
dimensions (with euclidean time, $X^0$,  compactified on a circle of
circumference $l$) transforms the  euclidean world-lines of
D-particles  to
D-instantons.   It is strongly believed that there are threshold
bound states
of  $n$ D-particles for any integer $n$.    The  configuration in
which the
world-line of such a state is  wrapped $m$ times has an  action
that  is  given
by,
\begin{equation}
\label{actiia}
\int_0^{2\pi} d t \left ( |n| e^{-\phi}  \sqrt {g_{00} } + i n (\alpha'
)^{-1/2} C^{(1)} {d X^0\over d t}\right),
 \end{equation}
 where  $C^{(1)} $ is the \RR one-form potential ($0\le C^{(1)} \le
2\pi$) and
$g_{00} = l^2(dX^0/dt)^2 (\alpha' )^{-1}$ (where $l$ is  the
circumference of
the compact euclidean time dimension).   For the $m$-wrapped
world-line  $
(\alpha' )^{-1/2}\int dt \ ({d X^0 / d t}) = 2\pi m$ and $\int dt
(g_{00})^{1/2} = 2\pi |m|l {\alpha'}^{-1/2}$.   Under T-duality
$C^{(1)}
\to C^{(0)}$ and $l e^{-\phi} {\alpha'}^{-1/2}  \to  e^{-\phi}  $.
Thus, the
action $S^{(mn)}$ for the  D-instanton with charge $p=mn$ is,
\begin{equation}
\label{manyinst}
 S^{(mn)}  = - 2\pi i (mn C^{(0)} + i |mn| e^{-\phi}) ,
\end{equation}
which is equal to $  -2\pi i|mn|   \tau$ for positive $mn$ (which
we shall
refer to as an instanton)  and $  2\pi i|mn| \bar \tau$ for negative $mn$
(which we shall refer to as an anti-instanton),  where $\tau =
C^{(0)} +
ie^{-\phi}$.\footnote{Thus, a D-particle wrapped in a positive
direction is
equivalent to an anti D-particle wrapped in a negative sense.}
Therefore
there are multiply-charged D-instantons of charge $p=mn$ with
degeneracy given
by the number of partitions of $p$ into two integers.    In section
6 we will
use this information, together with known results of string perturbation
theory, to motivate an $SL(2,Z)$ completion of the effective $R^4$
interaction
of section 5.    We will see that for this  to be correct there
must be an
interesting  non-renormalization theorem that forbids perturbative
corrections
to this term beyond one loop  in the zero-instanton sector -- a
property that
has a plausible heuristic origin.

In section 7 we will consider an intrinsically stringy effect that
arises from
a  world-sheet which is a disk   with  four massless closed-string tensor
states  attached.  The fixed-angle scattering of gravitons  on such a
world-sheet  is power
behaved  at high energy.  This  point-like behaviour arises, as in
the bosonic
theory,  from the presence of  massive closed-string states that
couple to the
boundary.  However, although this is  intriguing, its precise
interpretation is
unclear since  integration over fermionic modes causes  this
process to vanish
unless there are more external  particles that  are sources for the
 sixteen
fermionic moduli.    Furthermore,  we have not taken into account
possible
effects of instantons  with multiple charges that are required  by the
T-duality argument given above.

 Beyond lowest order in the coupling constant  scattering processes
generically
include connected world-sheets with more than one boundary, such as the
annulus with both boundaries fixed at the same space-time point.
This  is a
loop of  {\it closed} string -- a loop correction to the
D-instanton.   The
systematics of this perturbation theory involve cancellations between the
Dirichlet theory divergences that are
associated with the presence of  the bosonic translation zero  modes
\cite{polchinst,greengas}.  These will also be outlined in section
7.   The
boundary of moduli space that gives point-like scattering is not
contained
within  the region   that gives these cancelling divergences so that the
puzzling  point-like behaviour is not eliminated by cancellation of  the
Dirichlet divergences.

\section{Type IIB supergravity and the D-instanton}
The bosonic terms in the  low-energy IIB supergravity are
expressed  in a
manifestly $SL(2,R)$-invariant form in the Einstein frame,
\begin{equation}\label{iibact}
S={-1\over 2\kappa_0^2}\int d^{10}x \sqrt{-g} \left\{R - {1\over 2
\tau_2^2 }
  {\partial_\mu \tau \partial^\mu \bar \tau} -    {  1\over 12
\tau_2}(\tau
H_{NS} + H_R)_{\mu\nu\rho} (\bar \tau H_{NS} + H_R)^{\mu\nu\rho}
\right\}  +
\cdots ,
\end{equation}
 where the fifth-rank field strength has been set to zero and we
will set the
arbitrary constant coupling to the value $\kappa_0=1$ for
convenience.  The
terms represented by $\cdots$ in (\ref{iibact}) include kinetic
terms for
the spin-1/2 complex Weyl fermion, $\lambda$, and the complex Weyl
gravitino, $\psi_\mu$ (the 16-component spinor index is suppressed)
 and furthermore include an infinite series
of terms of higher order in derivatives and higher fermion terms.
 The scalar and antisymmetric tensor fields arise in (\ref{iibact})
in the
combinations,
\begin{equation}\label{complexconst}
\tau \equiv \tau_1 + i \tau_2 = C^{(0)}  + ie^{-\phi},\qquad
H=\left(\begin{array}{c}H_{NS}\\H_{R}\end{array}\right)  ,
\end{equation}
and $H_{NS} = dB_{NS}$, $H_R = d C^{(2)}$.
The scalar fields parameterize the coset space, $SL(2,R)/U(1)$, in
which the
$U(1)$ represents a local symmetry acting on the fermions.   Before
choosing a
gauge there are three scalar fields which enter the  zweibein
$V_{\alpha}^{\pm}$,
\begin{equation}
\label{sltwor}
V= {1\over \sqrt { -2i \tau_2}}\left(\begin{array}{cc}  \bar \tau
e^{-i\phi}
& \tau e^{i\phi}     \\
e^{-i\phi} & e^{i\phi} \end{array}\right),
\end{equation}
where $0\le \phi \le 2\pi$.  The local $U(1)$ rotations  act  from
the right so
that $V$ transforms into $VU(\alpha)$ (where $U= {\rm diag}(e^{i\alpha},
e^{-i\alpha})$)  which induces the shift,  $\phi \to \phi -
\alpha$.    The
global group $SL(2,R)$ acts from the left.
Following the notation in  \cite{schwarz} it is useful to define,
\begin{equation}
\label{pdef}
P_\mu \equiv - \epsilon_{\alpha\beta}V^\alpha_+ \partial_\mu V^\beta_+  =
ie^{2i\phi}{\partial_\mu \bar \tau \over 2 \tau_2} , \qquad Q_\mu
\equiv  - i
\epsilon_{\alpha\beta} V^\alpha_+ \partial _\mu V^\beta_-  =
\partial \phi -
{\partial_\mu \tau_1 \over 2\tau_2},
\end{equation}
where $Q_\mu$ is a composite $U(1)$ potential that couples
minimally to the
fermions.  The gravitino $\psi_\mu$ has $U(1)$  charge $1/2$ and
the dialtino
$\lambda$ has $U(1)$ charge $3/2$.   The definitions (\ref{sltwor}) and
(\ref{pdef}) correspond to those given in
\cite{schwarz}  after performing the $SL(2,C)$ transformation    that
takes the  IIB theory with the scalars living   $SU(1,1)/U(1)$
coset to the
one where the scalars parameterize a $SL(2,R)/U(1)$ coset.    The
gauge may be fixed by, for example, setting $\phi =0$, which   will be
used from here on.  This means that generic $SL(2,R)$ transformations on
the fields charged under $U(1)$ are associated with compensating
$U(1)$  transformations and the global symmetry is  nonlinearly realized.

Perturbation theory in  the zero-instanton sector is an expansion in
fluctuations around constant  values of the scalar fields, $\chi =
\langle
C^{(0)}\rangle$ and $\kappa = e^{-\langle \phi \rangle}$.
Defining  a complex coupling constant by,
\begin{equation}\label{coupdef}
\tau_0= \chi + {i\over \kappa},
\end{equation}
the fluctuating scalar fields can be expressed in a power series in
$\kappa$
as,
\begin{equation}
\label{fluct}
\tau \equiv C^{(0)} + i e^{-\phi} = C^{(0)} + {i \over \kappa} e^{-\kappa
\tilde \phi} = \tau_0 + \tilde C^{(0)}  - i\tilde \phi + {i\over 2}\kappa
\tilde \phi^2  + \cdots .
\end{equation}
where $\ \tilde{}\ $ denotes a quantum fluctuation of a field.

The $N=2$ supersymmetries are   $Q_I^A$  ($I=1,2$), where $Q_1^A$
and $Q_2^A$ are Majorana-Weyl spinors ($A=1,\cdots,16$) satisfying,
\begin{equation}\label{susyalg}
\left\{ \bar Q_I, Q_J\right\} = \delta_{IJ} \gamma\cdot p.
\end{equation}
In describing the instanton  we will consider the complex combinations
\begin{equation}
\label{plusmindef}
Q^\pm = \half (Q_1 \pm i Q_2).
\end{equation}
Defining $\theta^A = Q^{-A}$  it follows that  $Q^{+A} =  (\gamma\cdot p
\partial/\partial \theta)^A$.
The euclidean continuation of $Q^+$  annihilates the instanton
solution.  The
supersymmetry transformations of the fields can be found in
\cite{schwarz}.
 In particular, the fermion transformations have the form,
\begin{eqnarray}
  \delta \lambda =i\gamma^\mu
P_\mu\epsilon^*+..\label{ferm1}\label{lambdasusy} , \qquad\qquad \delta
\psi_\mu = D_\mu\epsilon+..\label{ferm2}
\label{psisusy}
\end{eqnarray}
where the spinor $\epsilon$ is a complex Weyl Grassmann variable so
that $\bar
\epsilon = \epsilon^* \gamma^0$.

The instanton is a euclidean saddle point of the bosonic part of
type IIB
supergravity  in which the two scalar fields have a nontrivial
profile (and
$H=0$).  The solution \cite{greenc} is one in which the
non-constant part of
the
 \RR\ scalar field  $C^{(0)}$ is imaginary so
that,\footnote{Classical values
of fields will
be denoted by a $ \hat{}$ in the following.}
\begin{equation}
\label{rrdist}
\hat \tau_1 = \hat C^{(0)}= \chi + i f(r),
\end{equation}
where $r= |x-y|$, $y^\mu$ is the position of the instanton, $\chi$
and $f$ are
real  and $f(r=\infty) =0$.
The fact that the instanton is a euclidean solution of the  BPS type  is
particularly clear if the theory is reexpressed in terms of an eight-form
potential $C^{(8)}$ by means of a duality transformation.   This
procedure,
which is implicit in \cite{greenc}, is described in detail in the
appendix.

As usual,  the  BPS  condition implies that  half of the euclidean
supersymmetries,
$Q^+$, annihilate the fields.  In other words, setting
$\epsilon^*=0$ in
(\ref{psisusy}) it turns out  that $\delta \lambda^* =0$ and
$\delta \psi =0$.
The first of these conditions follows directly from,
\begin{equation}\label{susyleft}
d\hat \tau_2 =  i d\hat \tau_1, \quad {\rm or} \quad  de^{-\hat
\phi} = -d f  .
\end{equation}
The instanton solution  has the form,    $\hat g_{
\mu\nu}=\eta_{\mu\nu}$
 and $\partial^2 e^{\hat \phi } =0$   for $x^\mu \ne  y^\mu$  so that
\begin{equation}\label{sol}
e^{\hat \phi } \equiv  h (r)  =  \kappa +{ c \over r^8},
\end{equation}
where  $\kappa = e^{\hat\phi(r=\infty)}$, and $c=3 |q| /\pi^{3/2}$ which
follows
from the quantization
condition  for a  D-instanton of charge $q$  and a euclidean seven-brane.
It will later be useful to represent $h$  in
momentum space by its Fourier transform,
\begin{equation}
\label{htrans}\tilde h(p)=\kappa \delta^{(10)}(p)+{c\over p^2}.
\end{equation}
Such an instanton has an action $S^{(q)} = 2\pi |q|/\kappa$, as
shown in the
appendix.

The solution can be transformed to the string frame by replacing the flat
Einstein-frame metric by the string metric, $g^{(s)}_{\mu\nu}=e^{\phi/2}
\eta_{\mu\nu}$.  In
the string frame the solution is a space-time Einstein--Rosen wormhole
\cite{greenc} which joins two universes that are related by the
interchange
$r\to (c /\kappa)^{ 1/4}/r$.   Since $e^{\hat \phi}$ gets large in
the neck
the field theoretic solution is not valid and the string theory
D-instanton
will have to  be used to account for short-distance physics.

Substituting the solution in (\ref{rrdist}) gives,
\begin{equation}
\label{soltau2}
\hat \tau_1 = \hat C^{(0)} = \chi + {i\over \kappa} - i e^{-\hat \phi} =
\tau_0 - i \hat \tau_2,
\end{equation}
so that
\begin{eqnarray}
\label{tauhatdef}
\bar {\hat \tau} = \tau_0 -   {2i \over  h},    \qquad   \hat \tau
=  \tau_0,
\end{eqnarray}
and $\hat \tau - \hat {\bar \tau}  = 2i/h$.

In order to satisfy the second  BPS condition,  $\delta \psi =0$,
the spinor
$\epsilon$  must be covariantly constant,  which implies,
\begin{equation}
\label{killing}
\epsilon =    j(r) \epsilon_0,
\end{equation}
where  $j(r) = (2/h)^{1/4}$ and  $\epsilon_0$ is a constant
sixteen-component
chiral spinor.   The broken
supersymmetries associated with the spinor $\epsilon^*$  will
generate the
instanton
solutions carrying fermionic zero  modes.

 The instanton   has the Minkowski-space interpretation   of a tunnelling
process in which  the initial  and final  \RR\ scalar
Noether charges differ by $q$ units, $q_f - q_i =q$, where
\begin{equation}\label{chargedefs}
q_i =\half  \int_{x^0 = -\infty}  d^9 x e^{2\hat \phi}  \partial_0 \hat
C^{(0)},
\qquad q_f= \half \int_{x^0 =  \infty}  d^9 x e^{2\hat \phi}
\partial_0 \hat
C^{(0)}.
\end{equation}
This uses the fact that the  Noether current associated with the shift
symmetry, $C^{(0)} \to C^{(0)} + b$, is   $j_\mu^N= e^{2\phi}
\partial_\mu
C^{(0)}/2$.    In general a charge $q$ instanton amplitude will be
accompanied by   a phase factor $e^{2\pi i
\chi q}$
where $0\le \chi \le 2\pi$ (as with the $\theta$ term in the case of
Yang--Mills instantons).  This is equivalent to introducing a
surface term in
the action,
\begin{equation}
\label{surfterm}
i \chi \oint_{\partial M_\infty}  *j^N = i\chi
\oint_{\partial_{M_\infty}}
\half e^{2\phi} * dC^{(0)} = i \chi \oint_{\partial M_\infty} \half F_9,
\end{equation}
which is an integral over the nine-sphere at $r=\infty$ (the last
expression
uses the dual potential described in the appendix).  Adding this
term to the
action  gives a total  action for a charge $q>0$ instanton  equal to
\begin{equation}\label{instact}
S^{(q)}  =  -2\pi i  |q| \tau_0 = 2\pi |q| ({1\over \kappa}  -i\chi),
\end{equation}
and for a charge $q< 0$ anti-instanton equal to
\begin{equation}\label{antiinstact}
S^{(q)}  =  2\pi i  |q| \bar\tau_0 =   2\pi |q|  ({1\over \kappa} +
i\chi).
\end{equation}

With $q=1$ this  is precisely the same as the action  obtained
 by expanding the
theory in
small fluctuations around the constant $\tau = \tau_0$ background in the
presence of a source term in the action,
\begin{equation}
\label{sourceact}
S_{source} = -2\pi i  \int  d^{10}x   \tau(x) \delta^{(10)} (x-y) =
 - 2\pi i
\tau_0.
\end{equation}
This is the D-instanton action that is the $p=-1$ case of the general
D-brane action.

The classical values of the fields that enter into the supergravity
are given
by
\begin{equation}
\hat P_\mu  =  - {\partial_\mu  h\over h}, \qquad \hat  Q_\mu =  -
{i \over 2}
{\partial_\mu  h\over h},  \qquad \hat P^*_\mu =0,
\end{equation}
so that  $\hat P^*$ is   {\it not} the complex conjugate of
 $\hat P^\mu$ since $\hat \tau_1$ is complex in the solution
 (\ref{tauhatdef}).
The momentum-space expression for $\hat P^\mu$ has a pole,
\begin{equation}
\label{momp}
\tilde{ \hat P^\mu} = {2i \over \kappa} {c p^\mu \over p^2} + \cdots,
\end{equation}
where the dots indicate non-pole terms.   Only the pole term, which
depends on
long-range effects,   will enter the lowest-order on-shell amplitude
calculations below.

As a simple example to illustrate the effect of the instanton
background we
will first consider the terms quadratic in the antisymmetric tensor
fields.  In
the vacuum of the zero instanton sector the scalar field is
constant,  $\tau =
\tau_0$,  and these terms have the form,
\begin{eqnarray}
S_{HH}^{(0)}=    {\kappa\over  24}\int d^{10}x   (\tau_0 H_{NS}+
H_R) (\bar
\tau_0 H_{NS} + H_R) = {\kappa \over 24}  \int d^{10}x G_0 \bar G_0,
\label{antiact}
\end{eqnarray}
where
\begin{equation}
\label{gdef}
G_0 = \tau_0  H_{NS} + H_R , \qquad  \bar G_0 = \bar \tau_0  H_{NS} + H_R
 \end{equation}
(these correspond to the $G$ and $\bar G$ defined in \cite{schwarz}
with the
scalar fields equal to constant vacuum values).

In the one-instanton sector the quadratic terms in the fluctuations
around the
one-instanton background are
obtained by substituting the expression for the instanton
configuration of the
scalar fields into the action (\ref{iibact}).  For example, for the
antisymmetric tensor fields the result of substituting the
classical solutions
$\hat C^{(0)}$ and $\hat \phi$ into (\ref{iibact}) is
\begin{equation}
\label{twotens}
S^{(1)}_{HH} = - 2\pi i   \tau_0 + S^{(0)}_{HH} + S'_{HH} + \cdots,
\end{equation}
where  $\cdots$ indicates higher-order terms and terms involving
fluctuations
of the scalar fields, and
\begin{eqnarray}
S'_{HH} &=&  {1\over 24}  \int d^{10}x    {c\over r^8}   ( \tau_0
H_{NS} +
H_R)  (\tau_0  H_{NS} +   H_R)\nonumber\\
& =&  {1\over 24} \int d^{10}x {c \over r^8} G_0   G_0.
\label{oneinst}
\end{eqnarray}
The action (\ref{twotens}) changes by $2\pi i \alpha$ under real
shifts  $\tau_0 \to \tau_0 + \alpha$  so that the effective action
will have the expected  factor of  $e^{2i\pi \tau_0}$ which is
invariant under integer shifts of the background $\hat C^{(0)}$.
The nonlocal
term (\ref{oneinst}) agrees precisely with the description of the
D-instanton
as the   source, (\ref{sourceact}),  which has a non-zero
contraction with
the  tree-level three-point couplings of two antisymmetric tensors.

\section{ Zero modes in a SUGRA   instanton background.}
The bosonic zero modes are parameterized by the  coordinates $y^\mu$
corresponding to the position of the D-instanton.  Integration over
$y^\mu$
enforces momentum conservation and results in a
factor of $1/\kappa^5$  (by a  very similar argument to the one
that results in
a factor
of $1/g^8$ in the measure for the zero mode integral in the
background of a
Yang--Mills instanton \cite{thooft}).

The fermionic zero modes can be determined, as usual, by applying
the broken
supersymmetry generators to the  scalar solutions.   The
transformations in
\cite{schwarz} can be adapted to the
present problem by making the identification $\epsilon^* = j(x)
\epsilon_0 $
(where $j(x)$ is defined in (\ref{killing})) and setting $\epsilon
=0$.   Since
$\epsilon_0$ has $16$ components this gives the possibility of $2^{16}$
independent separate instanton configurations obtained by applying
the broken
supersymmetries to the solution $\hat P_\mu$.  The following are
the terms
involving up to eight powers of $\epsilon^*$,
\begin{eqnarray}
 && \qquad \hat \lambda =  i \gamma^\mu \hat P_\mu   \epsilon^*,   \qquad
\hat G_{\mu\nu\rho}=3D_{[\mu}\bar{\epsilon}\gamma_{\nu\rho]}\hat
\lambda\nonumber\\
 && \qquad \hat  \psi_\mu={1\over 96}\left(
 \gamma_\mu^{\nu\rho\lambda}\hat
G_{\nu\rho\lambda}-9\gamma^{\rho\lambda}\hat
G_{\mu\rho\lambda}\right)\epsilon^*,  \qquad \hat
F^5_{\mu\nu\rho\lambda\sigma}= 5D_{[\mu}\bar{\epsilon}
\gamma_{\nu\rho\lambda}
\hat \psi_{\sigma]}\nonumber\\
&& \qquad  \hat e_\mu^r= -i \bar{\epsilon}\gamma^r\hat \psi_\mu,
\qquad\hat
\psi^*_\mu= -{1\over 480} i
 \gamma^{\rho_1\cdots\rho_5}\gamma_\mu
 \epsilon^*\hat F_{\rho_1\cdots\rho_5}\nonumber\\
 && \qquad \hat  G^*_{\mu\nu\rho}=12i
 \partial_{[\mu}\bar{\epsilon}\gamma_{\nu}\hat \psi^*_{\sigma]},
\qquad \hat
\lambda^*= {1\over 24}i \gamma^{\mu\nu\rho}\epsilon^* \hat
G^*_{\mu\nu\rho},
\qquad  \hat P^*_\mu = D_\mu \bar \epsilon^*   \hat
\lambda^*
\label{transferm}
\end{eqnarray}
(recall $\bar \epsilon = \epsilon^* \gamma^0$).  The successive terms are
defined iteratively in terms of $\hat P^\mu$.   These zero modes
are functions
with momentum-space poles that  can be  seen by using (\ref{momp}).

In making contact with string theory we will be interested in physical
fields satisfying the free equations of motion.  In that case it is
easy to see
that terms with more than eight  powers of $\epsilon^*$ in will not
be needed.
 This is particularly clear in a light-cone parameterization in
which the unphysical components of the fields are related to the
$2^8$ physical
components.   The physical
closed-string states can be packaged together into a light-cone scalar
superfield, $\Phi(x,\theta)$, where $\theta^a$ ($a=1,\cdots,8$) is an
eight-component $SO(8)$
spinor, ${\bf 8_s}$, (and the inequivalent $SO(8)$ spinor ${\bf
8_c}$ will be
represented by a dotted index). The equations of motion are imposed by
requiring $\partial^2 \Phi =0$.    With this notation a
16-component  chiral
spinor  has an $SO(8)$ decomposition,
\begin{equation}
\label{decomp}
\epsilon^{*A} \to (\eta^a,  \dot \eta^{\dot a})
\end{equation}
($A=1, \cdots,16$).
The broken supersymmetries are generated by $\eta^a Q^{-a} +\dot
\eta^{\dot a}
Q^{-\dot a}$, where $Q^{-a}$ and $Q^{-\dot a}$ are the $SO(8)$
components of
the broken supercharges and act on $\Phi$ by
\begin{equation}
\label{superact}
Q^{-a}  \Phi = \sqrt{p^+} \theta^a \Phi, \qquad Q^{-\dot a} \Phi =
\left({\gamma^i p^i \over \sqrt{p^+}}\theta\right)^{\dot a}  \Phi ,
\end{equation}
where $i=1,\cdots,8$ labels the directions transverse to the
light-cone.  The
classical fields, (\ref{transferm}), are simply fields satisfying (at the
linearised level)
\begin{equation}
\label{ttransferm}
\hat \Phi = (\eta^a Q^{-a} +\dot \eta^{\dot a} Q^{-\dot a}) \hat \Phi
\end{equation}
Since the components of $Q^-$ are linear in $\theta^a$ it is
evident that a
maximum of eight powers  can be applied to $\Phi$, resulting in a
maximum of
eight powers of $\epsilon$.

The equations (\ref{transferm}) define non-vanishing one-point
functions for
all components of a supermultiplet  in the background of a single
D-instanton.
The Grassmann parameters
are fermionic  supermoduli  corresponding to zero modes of $\lambda$ and
must be integrated over together with the translational zero modes,
$y^\mu$.
   A general Green function is given by an expression of the form,
\begin{eqnarray}
&& C\int d^{10} y d^{16} \epsilon_0  \langle   \Psi^1 (x_1) \Psi^2
(x_2)\cdots
\Psi^n (x_n)\rangle_{\epsilon_0,y} \nonumber\\
&& =
C \int d^{10} y d^{16} \epsilon_0 \int  D\Psi(x)
  \Psi^1 (x_1) \Psi^2 (x_2)\cdots
\Psi^n  (x_n) e^{-S^{(1)}},
\label{corrfuns}
\end{eqnarray}
where $\Psi^r$ represents any of the  fields of the theory
and $S^{1}$ is the action in the one-instanton background  which
depends
 on $y$ through its
dependence on  the background $\hat \tau$ field configuration.
We have not
determined the overall constant, $C$, which can depend on $\kappa$
and will
generalize to a function of $\tau$ and $\bar \tau$ when higher order
fluctuations are considered.  One way to pin down this dependence is to
understand how  the full effective action (including the sum over
all instanton
configurations) is invariant under $SL(2,Z)$.  This will be
discussed further
in section 6.

Since we will want to make comparisons with string theory  we will define
on-shell scattering amplitudes by the LSZ reduction that cancels
the poles on
the external legs in (\ref{corrfuns}).  In particular, the zero modes
 (\ref{transferm}) define on-shell
tadpoles in the presence of $s$ supermoduli,
\begin{equation}
 \label{LSZtrun}
\langle \Psi \rangle_s = \zeta_\Psi \Delta^{-1} \hat \Psi_s,
\end{equation}
where $\zeta_\Psi$ is the wave function for the on-shell
closed-string state
and $\Delta$ is its inverse propagator.

 The leading contribution to the amplitude that follows from
 (\ref{corrfuns}) is of the form
\begin{equation}
\label{leadinl}
e^{2\pi i  \tau_0} \int d^{10} y  d^{16} \epsilon_0 \langle \Psi
\rangle_{s_1}
\dots\langle \Psi
\rangle_{s_{n}},
\end{equation}
where $\sum_{i=1}^n s_i=16$.

\section{Zero modes in the  stringy D-instanton background. }
The simplest open-string world-sheet that arises in a D-brane
process is the
disk diagram.
 In the case of the D-instanton the  boundary satisfies Dirichlet
conditions in all ten space-time directions.  This means that
there are no
physical  propagating open strings but there is an
isolated open string supermultiplet consisting of a  vector
together with its
spinor superpartner (\cite{greeniib}
and references therein).   These remnants of the open-string sector
are the
zero-dimensional reduction of  the ten-dimensional massless  Yang--Mills
supermultiplet.
  The vector field corresponds to the collective coordinate given by  the
instanton position, $y^\mu$.  Any
world-sheet with vertex operators attached carrying momenta,
$k_i^\mu$ has an
overall factor of  $\exp(i\sqrt \kappa \sum_i k_i\cdot y)$ and
integration over
$y^\mu$ leads to momentum conservation.  The factors of $\sqrt
\kappa$ in the
open-string vertex lead to the factor of $\kappa^{-5}$ in the
measure for the
$y^\mu$ integration, as in the field theory.   The  open-string fermions
give rise to the   zero modes   in the  instanton background.
Integration
over the sixteen supermoduli leads to an overall  factor of
$\kappa^{ 8}$.  In
the complete string theory such factors of $\kappa$  will arise
from powers of
$\tau_2^{-1}$ and include the effects of dilaton couplings to the disk.

The disk with no states attached is defined by a functional integral that
is simply a constant and is identified with the source term in the
action,  (\ref{sourceact}), and is equal to the action for a single
D-instanton.

\medskip
\ifig\fone{An on-shell closed-string state, $\Psi$,   coupling to $s$
open-string fermions on a disk with Dirichlet boundary  conditions.  }
{\epsfbox{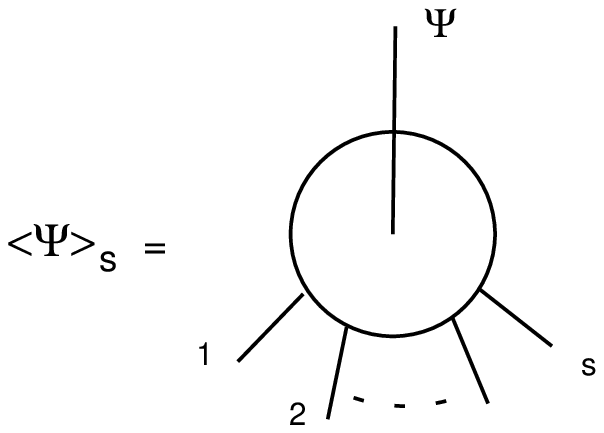}}
\medskip
An instanton carrying some zero  modes corresponds,  at lowest
order, to a disk
world-sheet with open-string states attached to the boundary (as in
\fone).
The
functional integration over the disk can be represented in an
operator approach
by transforming to a parameterization in which the world-sheet is a
semi-infinite cylinder with a boundary end-state, $|B\rangle$.
This state
was  constructed in  \cite{greeniib} to preserve one half of the
supersymmetries -- both the world-sheet supersymmetries and the
space-time
supersymmetries.   The formalism with manifest world-sheet supersymmetry
requires two kinds of boundary state -- $|B\rangle_{NSNS} $ and $
|B\rangle_{RR} $ -- in the \NSNS and \RR sectors, respectively.
These are
related by the requirement that the boundary state preserve space-time
supersymmetry.  This is manifest in the  formalism with manifest
space-time
supersymmetry  formulated in a light-cone frame defined by choosing
the axis of
the cylinder to be the light-cone time, $X^+$.
 In that case there is a single boundary state that is annihilated by  a
complex  combination of  left-moving and right-moving space-time
supercharges,
\begin{equation}
\label{presusy}
Q^+  |B\rangle \equiv (Q + i \tilde Q) |B\rangle =0.
\end{equation}
Enforcing this condition for all sixteen components of $Q^+$
determines the
boundary state, not just in the case of the D-instanton, but also
for all the
other $(p+1)$-instantons \cite{greengutperlec}.
The other combination, $Q^- = Q -i \tilde Q$, which  does not
annihilate the
boundary
state  is the broken supersymmetry.   This light-cone gauge formalism
should be derivable from a covariant supersymmetric formalism with
world-volume
 $\kappa$-symmetry, such as  in  \cite{nilsson,schwarzb,townsend}.

Applying a broken supersymmetry transformation  $s$  times defines
a state,
\begin{equation}
\label{multiferm}
|B\rangle_s \equiv \bar \epsilon_0^{A_1}Q^{-A_1} \cdots \bar
\epsilon_0^{A_s}Q^{-A_s}  |B\rangle,
\end{equation}
which  applies in either the  covariant formalism or the light-cone
formalism
In this notation a disk with  one physical on-shell closed-string state
attached to the interior  and $s$  fermionic open strings attached
to  the
boundary can be represented by
 \begin{eqnarray}
  \label{statecorre}
 \langle \Psi \rangle_s  = \langle \Psi | B\rangle_s = \langle
\delta^s \Psi |
B\rangle,
   \end{eqnarray}
where $\delta^s \Psi$ is the variation of $\Psi$ obtained by
applying $Q^-$ $s$
times.
Such tadpoles  should be equal to the LSZ truncated fields of the
supergravity
theory  (\ref{LSZtrun}).
This is confirmed by direct calculation of the disk amplitudes, as
follows.

In the light-cone gauge formalism the Grassmann spinor $\epsilon_0$
 is written
in terms of the  $SO(8)$ spinors $\eta^a$ and $\dot \eta^{\dot a}$.  The
closed-string tadpoles are then given  simply by the matrix elements,
\begin{equation}
\label{tadmat}
\langle \Psi \rangle_s = \langle \Psi |  \eta^{a_1}Q^{-a_1} \cdots
\eta^{a_s}Q^{-a_s}  |B\rangle\end{equation}
where each  index  $a_1,\dots, a_s$ may be  either undotted or dotted and
contracted into the  corresponding $SO(8)$ supercharge.  It is easy
to see that
the resulting expressions are precisely those obtained from the
field theory
after LSZ reduction.    The terms with purely undotted spinors
package together
into an $SO(8)$ superfield with components that are the $SO(8)$ physical
light-cone gauge components of the $\langle \Psi\rangle_s$ of
(\ref{LSZtrun}).
  The ground state
of this supermultiplet is the complex scalar, $(i
\zeta_\phi+\zeta_{C^{(0)}})$,
the linearized version of $\langle \tau \rangle$.  The next state
is $\zeta_\lambda^a \eta^a$, where $\zeta_\lambda^a$ is the complex  wave
function of the $SO(8)$
components of the spin-1/2 field, $\lambda^a = \lambda_1^a + i
\lambda_2^a$.
Continuing in this manner, all the components of the supermultiplet in
(\ref{transferm}) are reproduced.

These expressions can also be obtained covariantly by attaching  a single
closed-string  vertex operator to the interior of the disk (which will be
parameterized as the upper-half plane)  together with
fermionic open-string states attached to the boundary (the real axis).
 This may also be identified as the boundary state with $s$
fermionic open
strings  attached where the fermion vertex operator is given by,
 \begin{equation}
\label{fermhalf}
  F_{-{1\over2}}(x)=\bar \epsilon_0S(x)e^{-1/2\phi(x)}\;,\qquad F_{
  1\over 2}(x) = \bar \epsilon_0 \gamma_\mu S(x) \partial X^\mu(x)
e^{1/2\phi(x)},
\end{equation}
in the superghost number  $-1/2$   and $1/2$ pictures, respectively.
The components of $\epsilon_0$ are the 16-component spinor
supermoduli that
must  be integrated in
obtaining amplitudes.
The  total left-moving
 and  right-moving  superghost numbers on the disk must add up to
$-2$, which
can be
achieved by using combinations of vertices in appropriately chosen
pictures.

With an even
number of fermion open-string states attached to  boundary the disk
couples to
closed-string bosons.   The vertex operators in the $-2$ picture are
\begin{eqnarray}
  V^{NN}&=& \zeta^{NN}_{\mu\nu}\;e^{-\phi}\psi^\mu(z)
  e^{-\tilde{\phi}} \tilde{\psi}^\nu(\bar{z}) e^{ikX}\label{-1NSNS}\\
V_{(n)}^{RR} &=&   \zeta^{RR }_{(n)[\mu_1\cdots\mu_n]} e^{-\phi/2}
\bar S (z)
 \Gamma^{\mu_1\dots \mu_n}
\tilde {S}  (\bar{z}) e^{-3\tilde{\phi}/2} e^{ikX}.
\label{vertexdilaton}
\end{eqnarray}
Here, $\zeta^{NN}$ is the wave function for  the physical on-shell
states in
the \NSNS  sector, $\phi$, $B_{NS}$ or $G$.   The wave functions in
the \RR
sector,
$\zeta_{(n)}^{RR}$,  are $n$-form potentials describing $C^{(0)}$,
$C^{(2)}$
and
$C^{(4)}$.
With this choice of pictures an even number of    open-string operators
 can be coupled to the boundary of the disk with an equal number of
 $F_{+1/2}$ and $F_{-1/2}$ vertices.

The disk with an odd number of fermion states attached to the
boundary couples
to the  closed-string fermionic states  which are described by the vertex
operators in the $-3/2$ picture, \begin{eqnarray}\label{dilatino}
V^{RN}&=&
  \zeta^{RN}_{\mu a}      e^{-{\phi}/2}{S^a}({z_2})
e^{-\tilde{\phi}}\tilde{\psi}^\mu(\bar{z}_2)  e^{ikX}\\
\tilde V^{NR}   &=&    \tilde \zeta^{NR}_{\mu a}e^{-\phi}\psi^\mu(z_2)
  e^{-\tilde{\phi}/2}\tilde{S^a} (\bar{z_2})     e^{ikX} .
\end{eqnarray}
Evidently, in order to give a total superghost number of $-2$ there
must be  a
net superghost number of $-1/2$ from the open-string fermionic
states attached
to the
boundary.

 It is easy to evaluate the  expectation values of  a single
closed-string
vertex attached to the interior of a world-sheet  with  $s$
fermionic states
attached to the boundary  using standard   techniques.  Thus, for
$s=0$ the
only non-zero expectation
values arise for the combination $\langle i V_\phi + V_{C^{(0)}
}\rangle$,
which  is equal to $\langle \tau \rangle \sim  (i
\zeta_\phi+\zeta_{C^{(0)}})$
as before.

The zero modes of  the other fields in the supermultiplet are obtained by
attaching  fermionic open-string
vertex operators (\ref{fermhalf})   to the boundary.
  The expression for the disk with one fermionic open-string mode on the
boundary  and  one closed-string  fermion  attached to the interior
is given by
\begin{equation}
  \label{oneferm}
  \langle c F_{-{ 1\over 2}}(x)c\tilde{c}V^{RN} (z,\bar{z})\rangle.
\end{equation}
This is evaluated making use of the boundary condition that
reflects $\tilde S$
into $S$ and standard  product  expansions,  giving,
\begin{equation}
  \label{one-oneII}
\langle \lambda \rangle_1 =   (\zeta_{a\mu}^{RN   } + i
\zeta_{a\mu}^{NR})
\gamma^{\mu}_{ab}  \epsilon^b_0   e^{iky}  = \bar \zeta_\lambda
\epsilon_0
e^{iky} .
\end{equation}
Where $\zeta_\lambda^a= \gamma^\mu_{ab} (\zeta_{\mu}^{b\; RN   } + i
\zeta_{\mu}^{b\;NR})$ is the holomorphic combination of the two
dilatinos.
Attaching two fermionic zero modes to the
boundary likewise  gives non-vanishing one-point functions for a complex
combination of the \NSNS and \RR two-forms,   $B = B_{NS}+ i
C^{(2)}$.   This
comes from
the correlation function,
\begin{eqnarray}
\langle B \rangle_2 &=&  \langle c F_{ {1\over2}}(x_1)\int dx_2
F_{-{1\over2}}(x_2)c\tilde{c}  \left
(V^{NN} (z,\bar{z}) + i V_{(2)}^{RR} (z,\bar{z}\right) \rangle
\nonumber\\
&=&   \bar\epsilon_0\gamma^{\mu\nu\rho} \epsilon_0 k_{[\mu}
(\zeta^{NN}_{\nu\rho]} + i \zeta^{RR}_{\nu\rho]}) .
\label{nsnsast}
\end{eqnarray}

With three fermionic zero modes attached the expression for the
non-vanishing,
holomorphic gravitino tadpole is
\begin{eqnarray}
\langle \psi \rangle_3 &=& \langle c F_{ 1\over 2}(x) \int dx_2 dx_3
F_{-{1\over2}}(x_2)  F_{-{1\over2}}(x_3) c \bar c V^{RN} (z,\bar{z})
\rangle\nonumber\\
 &=&  \bar\epsilon_0 \gamma^{\mu\nu\rho}  \epsilon_0 \bar \zeta^{NR}_\rho
\gamma_{\mu}  \epsilon_0
k_\nu  .
\label{gravitinot}
\end{eqnarray}

 The disk with four fermion zero modes couples to the graviton and the
self dual fourth-rank antisymmetric tensor with are both invariant under
$SL(2,Z)$ S-duality of type IIB.   The choice of pictures for the
closed-string bosonic vertices in   (\ref{vertexdilaton})
requires that half
the open-string fermions  be in the $1/2$ picture and half in the
$-1/2$ so
that the tadpole is given by,
\begin{eqnarray}
\langle h \rangle_4 &=&  \langle cF_{1/2}(x_1)\int dx_2
F(x_2)_{1/2}\int dx_3
F_{-1/2}(x_3)\int
  dx_4 F_{-1/2}(x_4)c\tilde{c}  V^{NN} (z,\bar{z}) \rangle\nonumber\\
&=&  \bar\epsilon_0\gamma^{\rho\mu\tau }\epsilon_0 \;
 \bar\epsilon_0\gamma^{\lambda\nu\tau }\epsilon_0 \; \zeta_{\mu\nu}k_\rho
 k_\lambda ,
 \label{gravitonamp}
\end{eqnarray}
 which is the only covariant combination of four $\epsilon_0$'s,
two physical
momenta and the physical polarization tensor.

Continuing in this manner it  is clear that adding fermionic
open-string states
reconstructs the one-point functions of the field theory in
(\ref{transferm}).
All the terms with even numbers of fermionic moduli are   formed
from powers of
the matrix,
\begin{equation}
\label{genmod}
M^{\mu\nu} = \bar \epsilon_0 \gamma^{\mu\nu\rho} \epsilon_0 k_\rho,
\end{equation}
which is the most general bilinear in two spinors and linear in the
momentum.
The fermionic terms, such as (\ref{gravitinot}), have an  extra factor of
$\epsilon_0$.

\section{Lowest-order effective interactions  }

The one-instanton terms in the field theory  effective action  can
be deduced
by considering on-shell amplitudes in the instanton background.      The
integration over the fermionic moduli soak up  the sixteen  independent
fermionic zero modes.  The contributions that arise at  leading order in
$\kappa$    are of the form (\ref{leadinl})  which is a product of
the  \lq
classical' fields, $\hat \Psi^r$ in  (\ref{transferm}) with  a
total of sixteen
powers of $\epsilon_0$.   This is the analogue of the leading term in the
amplitude for gauge bosons in the Higgs-Yang--Mills instanton calculation
 in \cite{rubakov}, which is again determined by the classical
profile of the field.\footnote{We are very grateful to Steven Shenker for
pointing out this reference.}
In that case this particular kind of  contact interaction  is the
first term in
  a series that
reproduces an exponentially falling fixed-angle cross section with
a scale
that is symptomatic of the presence of solitonic states in the
theory.    An
analogous interpretation of the contact term in IIB supergravity is to be
expected.

The most obvious contact term is the one  proportional to
$\lambda^{16}$, which arises in IIB supergravity from the nonlocal
Green function, defining $r_m= \mid x_m-y\mid$
\begin{equation}
\label{greenfed}
G_{\lambda^{16}}(\{r_m\}) \sim  e^{2\pi i   \tau_0} \int d^{10} y  \int
d^{16}\epsilon_0 \prod_{m=1}^{16} \left( \gamma_\mu \hat P^\mu(r_m)
j(r_m)\epsilon_0\right) .
\end{equation}
where we have not kept track of overall  constant factors.
This  integral is well-defined because the function
$P^\mu(r_m)j(r_m)$ is
highly suppressed by the phase space volume at
the origin and well-behaved at infinity.
At long distances (\ref{greenfed}) looks like a $\lambda^{16}$
contact term.
This is the term that can be obtained by use of the LSZ procedure
described
earlier and leads to a  momentum-independent term in the S-matrix
with sixteen
external on-shell $\lambda$ particles proportional to,
 \begin{equation}
\label{lamcont}
e^{2\pi i  \tau_0} \epsilon^{A_1\cdots A_{16}} \lambda^{A_1}\dots
\lambda^{A_{16}}.
\end{equation}

 From the earlier discussion of the correspondence between the
string theory
and field theory zero modes it is evident that  the same result is also
obtained in string theory from diagrams with sixteen disconnected
disks with  a
single  $\lambda$ vertex operator  and a single open-string fermion state
attached to each one.   The overall factor of $e^{2\pi i  \tau_0}$,
which  is
characteristic of the stringy D-instanton \cite{polchinst,greengas}, is
evaluated at $\chi =\Re \tau_0 =0$ in the string calculation.  In a more
complete treatment the exponent should become $2\pi i  \tau$, which would
include interactions due to fluctuations of   $\phi$ and $C^{(0)}$.
 These
should be deduced in a systematic manner  by attaching vertex
operators for
these fields to further disconnected disks.   String theory also
contains
further diagrams that have no direct analogue in field theory, in
which more
than one closed-string state is attached to each disk.  Such
diagrams contain
the field theory contributions due to tree-level interactions in
the instanton
background but also contain intrinsically stringy effects, some of
which will
be described later.

We now turn to consider amplitudes with four external gravitons.
The leading
term in the field theory is again one in which each graviton is
associated with
four fermionic zero
modes, which gives a Green function,
\begin{equation}
\label{fourgrav}
G_{h^4} (\{r_m\})\sim e^{2\pi i  \tau_0}  \int d^{10} y  \int
d^{16}\epsilon_0
\prod_{r=1}^4  \left(\bar \epsilon_0
\gamma^{\mu_r\sigma_r\rho}\epsilon_0\ \bar \epsilon_0
\gamma^{\nu_r\tau_r}_{\rho}
\epsilon_0 \   k_r^{(\sigma_r}  \hat P_r^{\tau_r)} j^4(r_r) \right) .
\end{equation}

\medskip
 \ifig\ftwo{The leading contribution to the scattering of four
gravitons in the
D-instanton background.  The sixteen fermionic open strings represent the
supermoduli that must be integrated.  All disks have boundaries
fixed at the
same space-time point.}
{\epsfbox{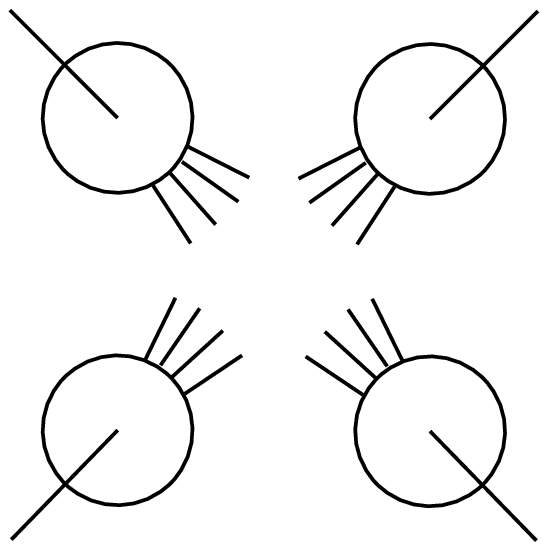}}
\medskip
Integration over $y^\mu$  generates a nonlocal four-graviton interaction.
Again, contact with the string calculation is made by considering
the on-shell
amplitude obtained by lopping off the external poles in momentum
space, making
use of
(\ref{momp}), and contracting the free indices with the on-shell wave
functions, $\zeta_h^{(r)}(k_r)$ which satisfy   $k_{r\mu}
\zeta^{(r) \mu\nu} =
k_{r \nu} \zeta^{(r) \mu\nu} = 0 = k_r^2$.    This corresponds to
the string
calculation   in which the world-sheet   (illustrated in \ftwo)
consists of
four disconnected disks to each of which is  attached a single
closed-string
graviton vertex and four fermionic open-string vertices.

In both the string theory and the field theory the  result is given as an
integral of the product of  four factors of $\langle h\rangle_4$
defined by
(\ref{gravitonamp}),
\begin{eqnarray}
\label{fourgravss}
A_4  (\{\zeta_h^{(r)} \}) &=& C \int d^{10} y  d^{16} \epsilon_0 \langle
h^{(1)}\rangle_4  \langle h^{(2)}\rangle_4  \langle
h^{(3)}\rangle_4  \langle
h^{(4)}\rangle_4 e^{i\sqrt \kappa k_r\cdot y}  \nonumber \\
&=&  C \int d^{10} y  d^{16} \epsilon_0 \prod_{r=1}^4 \left(  \bar
\epsilon_0
\gamma^{\mu_r\sigma_r\rho}\epsilon_0\ \bar \epsilon_0 \gamma^{\nu_r\tau_r
\rho}\epsilon_0 \zeta^{(\mu_r} \tilde \zeta^{\nu_r)} \
k_r^{\sigma_r}  k_r^{\tau_r}\;e^{i\sqrt{ \kappa} k_r\cdot y} \right),
\end{eqnarray}
where the polarization tensor have  been written as $\zeta^{\mu_r\nu_r} =
\zeta^{(\mu_r} \tilde \zeta^{\nu_r)}$, which is sufficiently
general for our
purposes (and overall constants, including possible factors of
$\kappa$, have
been absorbed into $C$).

 In order to evaluate the fermionic integrals we will choose a
special frame in
which $k_r^+ = k_r^- = 0$ and $k_r\cdot \zeta^{(r)} = k_r\cdot k_r =
\zeta^{(r)} \cdot \tilde \zeta^{(r)} =0$ (which is possible for
complexified
momenta and polarizations).
The spinor $\epsilon_0$ is naturally expressed in terms of its $SO(8)$
components  so that
\begin{eqnarray}
\langle h\rangle_4 &=&
\left(-\eta_{a}\gamma^{ij}_{ab}\eta_{b} \;\dot
\eta_{\dot{a}}\gamma^{mn}_{\dot{a}\dot{b}}\dot \eta_{\dot{b}}
+\eta_{a} \gamma^{kij}_{a\dot{b}}\dot \eta_{\dot{b}}\;
\dot \eta_{\dot{a}}\gamma^{kmn}_{\dot{a}{b}}\eta_{b}\right)
R_{ijmn}\nonumber
\\
&=&-{1\over 2}\eta_{a}\gamma^{ij}_{ab}\eta_{b} \;\dot
\eta_{\dot{a}}\gamma^{mn}_{\dot{a}\dot{b}}\dot \eta_{\dot{b}}  R_{ijmn}
\label{fourferms}
\end{eqnarray}
where $R_{ijmn} \equiv k_ik_m\zeta_{(j}\tilde \zeta_{n)}$ is the
linearized
curvature and the second line follows from the first after a Fierz
transformation.

 The integral over the dotted and undotted spinors in
(\ref{fourgravss}) now
factorizes and can be evaluated by using,
\begin{equation}
  \int d^8\eta^a \eta^{a_1}\cdots\eta^{a_8}=\epsilon^{a_1
  \cdots a_8}\;,\quad\qquad \;\int d^8\dot \eta^{\dot{a}}
\dot \eta^{\dot{a}_1}\cdots\dot \eta^{\dot{a}_8}=\epsilon^{\dot{a}_1
  \cdots \dot{a}_8}.
\end{equation}
Substituting in (\ref{fourgravss}) the following tensors appear
\begin{eqnarray}
  \epsilon_{a_1a_2\cdots a_8}\gamma^{i_1j_1}_{a_1a_2}\cdots
  \gamma^{i_4j_4}_{a_7a_8}&=&t^{i_1j_1\cdots
  i_4j_4}=\hat{t}^{i_1j_1\cdots i_4j_4}+{1\over2}\epsilon^{i_1j_1\cdots
j_4j_4}\label{t1}\\
  \epsilon_{\dot{a}_1\dot{a}_2\cdots
\dot{a}_8}\gamma^{i_1j_1}_{\dot{a}_1\dot{a}_2}\cdots
  \gamma^{i_4j_4}_{\dot{a}_7\dot{a}_8}&=&t^{i_1j_1\cdots
i_4j_4}=\hat{t}^{i_1j_1\cdots i_4j_4}-{1\over2}\epsilon^{i_1j_1\cdots
j_4j_4}\label{t2} ,
\end{eqnarray}
(where the notation follows \cite{tseytlinb}). There is an
important  sign
 difference between the eight-dimensional Levi--Cevita symbol  in
(\ref{t1})
and (\ref{t2}) which
can be seen by a careful decomposition of
 $SO(9,1)$ into the two inequivalent spinor representations ${\bf 8_c}$
 and ${\bf 8_s}$ of $SO(8)$. The result therefore contains two
parity-conserving
 terms
\begin{eqnarray}
A_4(\{\zeta^{(r)}_h\}) &=  & C  e^{2i\pi  \tau_0}   \int d^{10}y
e^{i \sum_r
k_r\cdot y}  \nonumber\\
&&\left(\hat{t}^{i_1j_1\cdots i_4j_4}\hat{t}_{m_1n_1\cdots
  m_4n_4}-{1\over 4}\epsilon^{i_1j_1\cdots j_4j_4}\epsilon_{m_1n_1\cdots
  m_4n_4}\right)  R_{i_1j_1}^{m_1n_1}
R_{i_2j_2}^{m_2n_2}R_{i_3j_3}^{m_3n_3}
R_{i_4j_4}^{m_4n_4} .
\label{termsin}
 \end{eqnarray}
It is notable that a parity-violating term,  proportional to  a single
$\epsilon^{i_1j_1\cdots j_4j_4}$ tensor does not contribute.    Although
obtained in a manner that is not
manifestly covariant these terms are easily covariantized by
extending the
indices to
the ten-dimensional range.   The term   bilinear in the eight-dimensional
Levi--Cevita tensor is the eight-dimensional Gauss--Bonnet term,
which vanishes
when the overall momentum is conserved.  Since momentum
conservation is only
imposed in the one-instanton sector after integration over $y^\mu$
this term is
non-zero for a fixed position of the instanton.

\section{An $SL(2,Z)$-invariant $R^4$ term and T-duality.}
The term  in
(\ref{termsin}) bilinear in the tensor $\hat t$  has precisely the
same form as
 terms
 that arise in the zero instanton sector that come both from the one-loop
four-graviton amplitude \cite{greenschwarziii} and from an
$(\alpha')^3$ effect
at
 tree level \cite{wittengross}.   Collecting these three  different
 contributions  together in the Einstein frame gives  an expression
 for the
complete  effective $R^4$ action that can be expressed in the form,
\begin{equation}
\label{rfourterms}
S_{R^4} = (\alpha')^{-1} \left[ a \zeta(3)\tau_2^{3/2}   + b
\tau_2^{-1/2} + c
e^{2\pi i  \tau} +\cdots \right] R^4 \equiv  (\alpha')^{-1}
f(\tau,\bar{\tau})R^4,
\end{equation}
where $R^4$ denotes the contractions $\hat t \hat t R^4$ in
(\ref{termsin}) and
$\cdots$ indicates possible perturbative and nonperturbative
corrections to the
coefficient of $R^4$.   The constants $a$ and $b$ given  in the
literature
depend on the normalization conventions (such as the value of
$\kappa_0$ in the
action (\ref{iibact})), although their dependence on $\tau_2$ is as
shown.  For
example,   in the normalization given in \cite{tseytlinb}, $a=
3\cdot 2^{-12}$
and $b= 3\cdot 2^{-18} \pi^{-5}$.    Since we have  not kept  track
of  the
factors of $\kappa$  the \lq constant' $c$ has not been determined
-- it could,
in principle, be a function of $\tau$ and $\bar \tau$.

However, the complete expression for $S_{R^4}$ must be invariant under
$SL(2,Z)$ transformations,  $\tau\to (a\tau+b)(c\tau+d)^{-1}$ (with
 integer
$a,b,c,d,$ satisfying $ad-bc=1$), which provides very strong
constraints on its
structure.
Since the $R^4$ factor is separately invariant the function
$f(\tau,\bar \tau)$
in (\ref{rfourterms}),  must transform as a scalar under the $SL(2,Z)$
transformations.
Such an expression necessarily involves a sum over  all instanton and
anti-instanton sectors.  Although we have not derived the form of
this sum
there are some strong constraints which this term must satisfy:
\begin{itemize}
 \item{} The tree-level perturbative contribution has the striking
coefficient
$ \zeta(3) = \sum_{m>0} (1/m)^3$.
\item{} The one-loop perturbative term is of order $\tau_2^{-2}$
relative to
this tree-level term.   The non-vanishing of this one-loop term is
possible
only  because the four external gravitons  are just sufficient to
soak up the
eight fermionic zero modes on a toroidal world-sheet.
Heuristically, the
perturbative $R^4$ terms beyond one loop should vanish since  extra
fermionic
zero modes on a higher genus surface  require additional  external
states to
give a non-vanishing  contribution.  Such a non-renormalization
theorem has not
appeared explicitly in the literature (as far as we know) and
deserves further
study.
\item{}  The multi-instanton contributions to the $R^4$ term can
only come from
a single  instanton   carrying multiple charge.  Separated
instantons carry
extra fermionic zero modes \cite{gutgreenus}  that lead to higher order
derivative interactions.
\item{}  Such  multiply-charged single  D-instanton configurations are
related by T-duality in nine euclidean dimensions to configurations
 in the
type IIA theory in which the world-line of  a single  D-particle
wraps $m$
times around the compact euclidean time dimension, $X^0$.   This
was shown  in
the introduction  where it was argued that these  instanton
contributions are
associated with a weight $e^{ - S^{(mn)}}  = e^{2\pi i |mn|
\tau}$ (and,
correspondingly,  $ e^{-2\pi i |mn|\bar \tau}$ for an anti-instanton).
Furthermore, T-duality on  a compactified $X^9$ direction relates the
D-particle to the D-string.   It then follows that these D-instanton
configurations can be related by T-duality in eight euclidean
dimensions to
configurations of  euclidean M-theory on $T^3$.
 \end{itemize}

Rather tantalizingly, there is a simple function that satisfies all these
criteria,  namely,
\begin{equation}\label{modularguess}
  f(\tau,\bar{\tau})= T_{pn}^{-3} \equiv \sum_{(p,n )\neq
(0,0)}{\tau_2^{3/2}\over |p+n\tau|^3},
\end{equation}
where $\sum_{(p,n )\neq (0,0)}$ indicates the sum is over all
positive and
negative values of $p,n$ except $p=n=0$.   It is easy to see that
this function
is invariant under $SL(2,Z)$ transformations.   The expansion of
this function
for small $\tau_2^{-1}$ is given by first  separating  terms in the
sum with
$n=0$ and representing the
rest as an integral,
\begin{equation}\label{firstf}
  f=2\zeta(3)\tau_2^{3/2} +{\tau_2^{3/2}\over \Gamma(3/2)}\sum_{n\neq 0,p }\int_0^{\infty}dy
y^{1/2}\exp\left\{-y(p+n\tau)(p+n\bar{\tau})\right\}
\end{equation}
Now the sum over $p$ should be reexpressed using  the   Poisson
resummation
formula,
\begin{equation}
  \sum_{p=-\infty}^{\infty}\exp(-\pi A
  (p+x)^2)=A^{-1/2}\sum_{m=-\infty}^{\infty}\exp\left (-{\pi
m^2\over A}+2\pi
  i m x\right),
\end{equation}
which gives,
\begin{equation}\label{poisonresummed}
   f=2\zeta(3)\tau_2^{3/2} + {2\pi^{2}\over 3}\tau_2^{-1/2} +
2\tau_2^{3/2}\sum_{m,n\neq 0}\int_0^{\infty}dy
  \exp\left(-{\pi^2 m^2\over y}+2\pi i m n\tau_1 - yn^2 \tau_2^2
\right),
\end{equation}
 where the second term accounts for the $m =0$ terms and we have
used $\sum_n
n^{-2}=\pi^2/6$. The rest of the integral can either be evaluated
by using a
saddle point method
or simply related to  a  $K_{1}$ Bessel function so that,
\begin{eqnarray}
   f (\tau, \bar \tau)&=& 2\zeta(3)\tau_2^{3/2} +
  {2\pi^2\over
3} \tau_2^{-1/2}  +8 \pi  \tau_2^{ 1/2}  \sum_{m \ne 0 n\ge 1   }
\left|{m\over n}\right|
e^{2\pi i mn\tau_1} K_1 (2\pi |mn|\tau_2)\nonumber\\
&=& 2\zeta(3)\tau_2^{3/2} + {2\pi^2\over  3} \tau_2^{-1/2}   \nonumber\\
&& +4\pi^{3/2}  \sum_{m,n \ge 1} \left({m\over n^3}\right)^{1/2}
(e^{2\pi i mn
\tau} + e^{-2\pi i mn \bar  \tau} ) \left(1 + \sum_{k=1}^\infty  (4\pi mn
\tau_2)^{-k} {\Gamma(  k -1/2)\over \Gamma(- k -1/2) } \right) ,
 \label{bessum}
\end{eqnarray}
where we have used the asymptotic expansion for $K_1(z)$ for
large $z$ in
the second equation.

This expression  incorporates all the points itemized earlier.  It is an
expansion with the appropriate perturbative terms (whose relative
normalization
differs from that in \cite{tseytlinb}  where the definition of
$\tau_2$ was
different).  The perturbative terms terminate after the one-loop term as
suggested earlier.  The non-perturbative terms have the form of a
sum over
single multiply-charged instantons and anti-instantons with action
proportional
to $|mn|$.  If the original integer $p$ in (\ref{modularguess}) is
identified
with the discrete momentum (euclidean energy)  of a compactified
D-particle of
charge $n$ then the Poisson resummation exchanges it for the
winding number of
the world-line and the result is that  expected by T-duality from
type IIA in
nine dimensions.  The terms in parenthesis in (\ref{bessum})
represent the
infinite sequence of  perturbative corrections around the
instantons of  charge
$mn$.   Such corrections, beginning with the $\tau_2$-independent
term, ought
to be obtainable directly from the D-instanton calculation which
would give  an
important check of the form of $f$.   Thus, although we have by no
means proven
that   the conjectured expression for $f$ in (\ref{modularguess}) is the
$SL(2,Z)$-invariant  coefficient of the $R^4$ term it satisfies several
stringent constraints.

In pursuing possible connections with M-theory it should be of
significance
that the function $f$  is the $s=3/2$ case of an expression of the form
\begin{equation}
\label{eigenvals}
f_s(\tau,\bar{\tau})= \zeta_\Delta (s) = \sum_{(p,n)\ne(0, 0)} (\lambda_{p,n})^{-s}
\end{equation}
where $\lambda_{p,n}$ are the eigenvalues of the laplacian on a
two-torus.
Explicitly,
the laplacian on a two-torus with modular parameters $\tau_1$ and
$\tau_2$  has
eigenfunctions
\begin{equation}
  \label{eigenf}
  \psi_{p,n}(x,y)=\exp\left\{ 2\pi inx+ 2\pi iy({p\over \tau_2} +
{n\tau_1\over
\tau_2})\right\},
\end{equation}
which satisfy the eigenvalue equation,
\begin{equation}
  \Delta\psi_{p,n}=
(\partial_x^2+\partial_y^2)\psi_{p,n}
=-{4\pi^2\over \tau_2^2}|p + n\tau|^2\psi_{p,n} \equiv \lambda_{p,n}
\psi_{p,n}.
\end{equation}
 The integers, $p$ and $n$, are Kaluza--Klein momenta which are
interpreted as
the discrete energy and charge of a $D$-particle compactified to nine
euclidean dimensions.
 The function $f_s$ is a generalized Eisenstein series defined in
\cite{zagiera,tarrasa}\footnote{We are grateful to Greg Moore and Richard
Borcherds for pointing these references out to us.} by
\begin{eqnarray}
E^*(\tau,\bar{\tau},s) &=& \half \pi^{-s} \Gamma(s) f_s(\tau,\bar{\tau})\\
&=&\half \pi^{-s} \Gamma(s) \sum_{(p,n)\ne
(0,0)}\left({\tau_2
\over |p + n\tau |^2}\right)^s\nonumber \\
   &=& \pi^{-s} \Gamma(s)  \zeta(2s) E(\tau,\bar{\tau},s),
\label{staredef}
\end{eqnarray}
where  $\zeta(s)$ is Riemann's zeta function,
\begin{equation}
\label{edef}
E(\tau,\bar{\tau},s)  =  \half   \sum_{  (l_1|l_2) =1}\left({ \tau_2 \over
|l_1\tau +
l_2|^2}\right)^s,
\end{equation}
 and $l_1$ and $l_2$ are relatively
prime.  One possibly significant property of these functions is that they
satisfy the eigenvalue equation,
\begin{equation}
\label{eeigen}
 \tau_2^2 (\partial_{\tau_1}^2 +
\partial_{\tau_2}^2)E^*(\tau,\bar{\tau},s) = s(s-1) E^*(\tau,\bar{\tau},s).
\end{equation}

\section{ Fixed-angle scattering.}
The leading contribution to the  scattering  amplitude  for four
closed-string
states in the one-instanton sector  is the term considered above in
which each
external graviton is
attached to a separate disk (\ftwo), thereby generating   the $R^4$
contact
term.  As we saw,
the presence of this term can also be deduced within type
IIB supergravity and there is nothing intrinsically stringy about
it.  This
contact term   grows like a positive
power  of  the energy and should not be interpreted as   a sign of
fundamental
point-like structure.  Such terms arise also in   the analogous
calculation of
high energy scattering in the presence of Yang--Mills instantons
\cite{rubakov}.  There they are   the leading terms term in a
power series expansion of a function that decreases exponentially
with energy
at fixed angle.

In this section we will consider higher-order processes, in which
two or more
closed-string states are attached to a disk but, for simplicity, we
will not
attach fermionic open-string states  to  the boundary.  Such
processes would
obviously vanish after integration over the fermionic modes unless
there are
other external states to act as sources for the sixteen fermionic
moduli, so
they should really be considered to be  sub-processes in
amplitudes with more
external states  (the analogues of  terms in the action in a
one-instanton
background).   We will  here simply evaluate such  diagrams without
performing
the  integrations over the fermionic moduli.

 Diagrams in which there are two  vertex operators attached to a
single disk
lead to contributions  that  are of order $\kappa$.   For example,
the coupling
of two massless tensor states   to a single disk with no external
fermions is
easily evaluated by considering $\langle V^{NN} V^{NN} \rangle$, $\langle
V^{RR} V^{RR} \rangle$ and $\langle V^{NN} V^{RR} \rangle$.  These
quantities
are readily  evaluated and the result is that the two-graviton
diagram vanishes
\cite{klebanova} (as it does in the bosonic theory
\cite{greengas}).  However,
   the process  with two antisymmetric tensors is precisely the
same as that
deduced from the   action (\ref{oneinst}).

\medskip
\ifig\fthree{Four closed-string states attached to a disk with Dirichlet
boundary conditions. Since there are no fermionic moduli this
diagram is a
sub-process in a complete amplitude.}
{\epsfbox{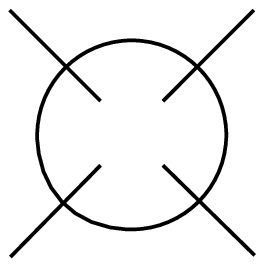}}
\medskip
The diagram in \fthree,  is the lowest-order diagram with  four
gravitons  that
has  no open-string states attached.     Even though this only
gives a non-zero
contribution to the S-matrix in the presence of sources for the fermionic
moduli  it is   instructive to analyze its behaviour in isolation
since it
illustrates explicitly the point-like  effects of   Dirichlet boundary
conditions.    Furthermore,  with general external states this
diagram also
contains the open-string divergences characteristic of Dirichlet boundary
conditions.  These arise when a bosonic intermediate open string
degenerates
\cite{greendual}   and  are guaranteed to cancel when all diagrams
of a given
order  in $\kappa$ -- including disconnected diagrams -- are added
correctly
\cite{polchinst}, as we will review below.

In evaluating the diagram in \fthree\   there are many equivalent ways of
assigning ghost numbers to the
vertices in order to ensure  that the total superghost number on the
disk is $-2$ but it will be simplest to choose all the vertices to
be in the
$(0,0)$ picture apart from the vertex at $z_1=i$, which will be
taken to be in
the $(-1,-1)$ picture.\footnote{In fact,  it is interesting to see
how the
  picture changing symmetry is preserved in this situation by the
  presence of boundary terms that are required in order  to ensure
  the antisymmetric tensor gauge invariance \cite{gutthesis}.}
The amplitude
is given by,
  \begin{equation}
\label{fourgravscat}
A_{4h}  = \int d y_2 d^2 z_3 d^2 z_4 \left \langle
c\bar{c}V^{NN}(\zeta^{(1)},z_1)(c+\bar{c})
V^{NN}_{(0,0)}(\zeta^{(2)}, iy_2)
V^{NN}_{(0,0)}(\zeta^{(3)} , z_3) V^{NN}_{(0,0)}(\zeta^{(4)} ,
z_4)\right\rangle,
\end{equation}
where the coordinates of the vertices, $z_r = x_r+ i y_r$,  span
the upper-half
plane and we have fixed the   M\"obius symmetry by locating one vertex at
$z_1 =i$ and a second on the imaginary  axis so that $z_2=iy_2$
($0\le y\le
1$).
The vertex in the $(-1,-1)$ picture is given by (\ref{-1NSNS})
 and the vertex in the $(0,0)$ picture is given by
\begin{equation}
   \label{0NS}
    V^{NN}_{(0,0)}=\zeta_{\mu\nu}(\partial
X^\mu+ik_\rho\psi^\rho\psi^\mu)(\bar{\partial}X^\nu+
ik_\lambda\bar{\psi}^\lambda\bar{\psi}^\nu)e^{ikX(z)}
 \end{equation}

Recall that we are considering   this diagram to be a sub-process
in which
other external
 physical states may provide a total momentum   $q$ (which could also
 be taken
to vanish).  After  integrating over the position of the instanton
 the total
momentum is conserved, $\sum_{r=1}^4 k^\mu_r + q^\mu =0$, so that
$s+t+u +
q^2=0$, where  $s=-(k_1+k_2)^2$, $t= -(k_1+k_4)^2$ and
$u=(k_1+k_3)^2$.
 We shall consider a limit in which $s$, $t$ and $u$ are all large,
which includes
the \lq fixed-angle' limit in which $q^2$ is small  and
\begin{equation}\label{kinematics}
s\to \infty, \qquad  t\to -\infty, \qquad  {t\over s} =
{1\over 2} (\cos\theta
-1),
\end{equation}
where  the  scattering
angle, $\theta$, is fixed.      In open-string theories with Neumann
boundary conditions as well as conventional closed-string theories the
behaviour
of the string
amplitude in this limit  is extremely soft
 \cite{veneziano,amati,grossmende}
--
the amplitude  falls off  exponentially as a function of $s$,
$A(s,\theta)
\sim \exp(-Cs)$ where the  constant $C$ depends on $\theta$ and on the
genus of the world-sheet but not on the details of the
string theory considered.  This stringy behavior is also
characteristic of scattering from D-branes  with $p\geq 0$ in
perturbation
theory \cite{klebanova,barbon} where the process is dominated by
the cloud of
open
strings surrounding the brane.   The presence of  a Dirichlet boundary
condition in the time direction leads to point-like scattering in
the bosonic
theory \cite{greenpointlike}.   We will see that this is also true
for the
D-instanton of the
type IIB superstring.   The same is true for any of the
$(p+1)$-instantons  no
matter how many compactified Neumann directions there are.

The amplitude  (\ref{fourgravscat})   has an
integrand that is the product of  an assortment of  prefactors   that
multiply a universal  momentum-dependent factor,
\begin{equation}\label{momentum}
\exp(E)=
\exp\left\{ \sum_{i\neq j}k^i \cdot k^j\ln\left|
\frac{z_i-z_j}{z_i-\bar{z}_j}\right|\right\},
\end{equation}
that depends on the scalar  Green function on the half plane,
with  coordinates satisfying  Dirichlet boundary conditions.  The
fact that
amplitudes decrease as a power of the invariants in the asymptotic
limits such
as (\ref{kinematics}) is determined by the form of (\ref{momentum})
although
the prefactors will need to be considered in order to determine the
precise
power.   The factor  $e^E$ is   exponentially suppressed  at high
energy and
fixed angle
unless the all but one of the $z_i$ approach  the boundary of the
disk.  In
that case the exponent vanishes.   Using M\"obius symmetry to fix
one vertex at
the origin this means that   high-energy fixed-angle behaviour  is
dominated by
the boundary  of moduli space in which the other three vertex operators
approach the boundary.
In the case of Neumann boundary conditions the exponent  in
(\ref{momentum}) is
replaced by $\sum_{i\neq j}k^i\cdot k^j\ln \left|
{(z_i-z_j)}{(z_i-\bar{z}_j)}
\right|$ which never vanishes so that the amplitude is always
exponentially
suppressed.

The relevant
boundary of moduli space that dominates in the fixed-angle limit is
$y_i\to 0$.
  Expanding  the terms in (\ref{momentum})  around $y_i= 0$ gives
\begin{eqnarray}\label{limit1}
 \ln\left|\frac{z_1-z_i} {z_1-\bar{z_i}}\right|^2&=&-\frac{4y_i}
{1+x_i^2}+O(y_i^2);\qquad
i=2,3,4\\
\ln\left|\frac{z_i-z_j}{z_i-\bar{z_j}}\right|^2&=& O(y_i^2);\qquad
i,j=2,3,4,\label{limit2}
\end{eqnarray}
where it has been assumed that $y_i\ll |x_i -x_j|$.
Hence the factor $e^E$  behaves  as
\begin{equation}\label{limit3}
\exp\left(E(s,t,q^2; y_2,y_3,y_4,x_3,x_4) \right)\sim \exp\{4sy_2+
\frac{4u}{x_3^2+1}y_3+ \frac{4t}{x_4^2+1}y_4+O (y_i^2)\}.
\end{equation}

It is easiest to analyse the asymptotic behaviour   in the \lq deep
euclidean' limit in which  $s,t$ and $u\to -\infty$ (so that $q^2 \to
+ \infty$)  since the exponent is positive definite in
this kinematical region.  In that case the integral is dominated   by the
region in which $y_2 \sim s^{-1}$, $y_3\sim u^{-1}$ and $y_4 \sim
t^{-1}$.
Thus, integrating with constant measure   gives a behaviour $\sim
(stu)^{-1}$.

This power behaviour should also be recovered in any limit with
large $|s|$,
$|t|$ and $|u|$.  These more general asymptotic limits can be analyzed
explicitly by
analytically continuing the $s,t,u$ invariants from a region in which the
expression (\ref{limit3}) is
well-defined.   We shall consider the example of the  limit in
(\ref{kinematics}) where $q^2$ is fixed.  It is convenient to use
the conformal
transformation,
\begin{equation}\label{map}
e^{\tau+i\sigma}=\frac{i-z}{i+z},
\end{equation}
to map the upper-half plane onto the semi-infinite cylinder with the
boundary at $\tau =0$.
The  factor (\ref{momentum}) reduces to
\begin{equation}\label{limit4}
\exp\{-s\tau_2-t\tau_3-u\tau_4+ O (\tau^2)\},
\end{equation}
where $-\infty<\tau_i<0$.    If we now continue  $s,t\to -\infty$
(implicitly
letting $u\to  +\infty$)  the exponent becomes
\begin{equation}\label{limit5}
\exp\{ -s(\tau_2-\tau_4)-t(\tau_3-\tau_4)+q^2\tau_4+  O (\tau^2)\},
\end{equation}
which is convergent if $\tau_2<\tau_4$ and
$\tau_3<\tau_3$.   Since $\tau$ corresponds to the proper time along the
cylinder  this picks out a specific ordering of
the three vertex operators along the cylinder, where $\tau_4$ is
closest to the
boundary.   If  the integration over $\tau_i$ is carried out
ignoring the fact
that there are other prefactors in the full expression for the
amplitude this
ordering gives a contribution of $(stq^2)^{-1}$. The higher terms in the
expansion are exponentially suppressed
in this limit.    For other $\tau$ orderings the integral is defined by
taking other limits of $s,t$ and $u$.   These are the orderings in
which the
vertex operators 2  and  3 are
closest to the boundary which corresponds to $u,t\to-\infty$ (with $s\to
\infty$)  and
$s,u\to -\infty$  (with $t\to \infty$), respectively.  Adding all three
contributions
together (still ignoring the effect of prefactors in the integrand)
gives a
total factor proportional to
\begin{equation}\label{stu}
\frac{1}{q^2}(\frac{1}{st}+\frac{1}{tu}+\frac{1}{su})=\frac{s+t+u}{q^2stu}=
-
\frac{1}{stu}.
\end{equation}
As anticipated, this is the same asymptotic behaviour as in the
limit $s,t,u \to -\infty$.

This is the essence of the analysis of the fixed-angle behaviour but  the
prefactors in (\ref{fourgravscat})  contribute powers of the
invariants which
depend on
which particular particles are scattered and upon their helicity
states.   To
begin with we will  consider terms in the amplitude that involve
contractions
between all the external polarization tensors in a cyclic order,
i.e., terms
with prefactors  of the form $\zeta^{(1)}_{\mu_1\mu_2}
\zeta^{(2)\mu_2\mu_3}
\zeta^{(3)}_{\mu_3\mu_4} \zeta^{(4)\mu_4\mu_1}$.    It has
previously been
argued \cite{greendual,gutperlea}  that such terms are particularly
simple
because the
divergences associated with degenerations of intermediate open-string
world-sheets do not contribute.  Terms of this form will be
referred to as  \lq
cyclic'
contractions.
Obviously, other kinds of terms, such as those in which  momenta
contract into
the polarization tensors arise in the complete process and will  be
considered
later.

\subsection{ Cyclic contractions.}

In addition to the fact that there are three inequivalent cyclic
contractions
corresponding to the cyclically inequivalent permutations of the external
states each of these  consists of sixteen different terms since each
polarization tensor can be contracted into its neighbour on either of two
indices.  Writing $\zeta^{(r)}_{\mu\nu} = \zeta^{(r)}_\mu \tilde
\zeta^{(r)}_{\nu}$  the  terms contributing to the $(1,2,3,4)$ cycle have
kinematic prefactors of the form,
\begin{eqnarray}\label{termss}
&& K(1,\bar 2; 2, \bar 3; 3, \bar 4; 4, \bar 1) =  \zeta^{( 1
)}\cdot \tilde
\zeta^{(2)} \ \zeta^{( 2 )}\cdot \tilde \zeta^{( 3 )}\  \zeta^{( 3
)} \cdot
\tilde\zeta^{(4 )}\  \zeta^{( 4 )}\cdot \tilde \zeta^{( 1 )},\nonumber \\
&&  K(1, 2;\bar 2, \bar 3; 3,   4; \bar 4, \bar 1) = \zeta^{( 1 )}\cdot
\zeta^{(2)} \  \tilde\zeta^{( 2 )}\cdot \tilde \zeta^{( 3 )}\
\zeta^{( 3 )}
\cdot  \zeta^{(4 )}\  \tilde\zeta^{( 4 )}\cdot \tilde \zeta^{( 1
)},, \nonumber
\\
&&  K(1,\bar 2; 2,   3;\bar 3, \bar 4; 4, \bar 1) =  \zeta^{( 1
)}\cdot \tilde
\zeta^{(2)} \ \zeta^{( 2 )}\cdot  \zeta^{( 3 )}\  \tilde \zeta^{( 3
)} \cdot
\tilde\zeta^{(4 )}\  \zeta^{( 4 )}\cdot \tilde \zeta^{( 1 )},
\nonumber \\
&&\qquad\qquad\qquad\quad\quad \cdots \cdots
\end{eqnarray}
If the external wave functions have no particular symmetry each of these
prefactors multiplies an independent amplitude.  However, if the
$r$th external
particle is a graviton then the amplitude is symmetric under the
interchange $r
\leftrightarrow \bar r$ while the amplitude is antisymmetric under this
interchange if the $r$th particle is an antisymmetric tensor state.
    The resulting term in the amplitude with  a prefactor of
the form of the first line in (\ref{termss}) is proportional to
\begin{equation}\label{fourgravscat2}
\int dy_2 d^2 z_3 d^2 z_4  { K(1,\bar 2; 2, \bar 3;3, \bar 4; 4, \bar 1)
k_2.k_4  (1- y_2^2) \over (iy_2+z_4) (i+iy_2)(iy_2-\bar{z}_3)^2
(z_3-\bar{z}_4)^2  (z_4+i)} e^E.
\end{equation}

The presence of   inverse powers of world-sheet coordinates in the
integrand
alters the asymptotic fixed-angle behaviour that was deduced from
the integral
of   $e^E$ with constant measure.    The integral is still
dominated by the
end-point of the integration in which three vertex operators approach the
boundary: $y_2,y_3,y_4 \sim 0$.   In this limit the denominators in
(\ref{fourgravscat2}) are finite at generic values of $x_3$ and
$x_4$ and the
power
behaviour of the amplitude is enhanced  from (\ref{stu})  to
$(st)^{-1}$ by
the factor of $k_2\cdot k_4 = - u/2$ in the numerator.   The
end-points in
moduli space at which two vertices approach each other on the world-sheet
boundary  (such as $z_3 \to \bar z_4$)  require special
consideration because
these are the regions in which the denominators in (\ref{fourgravscat2})
vanish.  It
is precisely these boundaries of the integration region that can
give rise to
the novel Dirichlet logarithmic divergences illustrated in
\cite{polchinst,greengas}  which should cancel with corresponding
divergences
arising from the degeneration limit of the annulus.      However,
these divergences only couple in channels with scalar quantum
numbers so they
do not arise for  the parts of the amplitude involving cyclic
contractions,
which is the main reason for choosing  to analyze  these terms first.

 If all the external states are gravitons the Dirichlet divergences
 are, in
fact,  absent
from  the non-cyclic, as well as the cyclic,  contractions.  The
behaviour of
the full amplitude can be
deduced from the term (\ref{fourgravscat2}) together with Bose
symmetrization
and
gauge invariance.  For this reason  the four-graviton amplitude will be
considered first.

\vskip 0.3cm
\noindent{\it Four external gravitons}\hfill\break
The amplitude with external gravitons is obtained by identifying $\tilde
\zeta^{(r)}$ with $\zeta^{(r)}$ so that all sixteen  terms for the
ordering
$(1,2,3,4)$ must be added with equal weight.   This symmetrization
leads to
cancellations so that the leading power behaviour of the process is
suppressed.

These cancellations are well illustrated by comparing the contractions,
\begin{equation}\label{exch1}
\zeta^{(1)}_{ \mu_1}\zeta^{(2)}_{ \mu_2} \tilde \zeta^{(2)}_{
\nu_2} \tilde
\zeta^{(3)}_{ \nu_3} k^2_{\rho_2} k^4_{  \rho_4} \langle
\psi^{\mu_1}(z_1)
\tilde{\psi}^{\nu_2}(\tilde{z}_2)\rangle\langle
\partial
X^{\mu_2}(z_2)\bar{\partial}X^{\nu_3}(\bar{z}_3)\rangle
\langle \tilde{\psi}^{\rho_2}(\bar{z}_2)
{\psi}^{\rho_4}(z_4)\rangle,
\end{equation}
that arise in the calculation of the amplitude with the factor
obtained by
interchanging $\zeta^{(2)}_{\mu_2}$ with $\tilde
\zeta^{(2)}_{\nu_2}$ and $z_2$
with $\bar z_2$
\begin{equation}\label{exch2}
\zeta^{(1)}_{\mu_1}\zeta^{(2)}_{\mu_2} \tilde \zeta^{(2)}_{\nu_2} \tilde
\zeta^{(3)}_{ \nu_3} k^2_{\rho_2} k^4_{ \rho_4} \langle \psi^{\mu_1}(z_1)
{\psi}^{\mu_2}(z_2)\rangle\langle
\bar{\partial}
X^{\nu_2}(\bar{z}_2)\bar{\partial}X^{\nu_3}(\bar{z}_3)\rangle
\langle {\psi}^{\rho_2}({z}_2)
{\psi}^{\rho_4}(z_4)\rangle.
\end{equation}
With  $\tilde \zeta^{(2)} = \zeta^{(2)}$ then both these terms
contribute a
factor
of   $k^2\cdot k^4\ \zeta^{(1)} \cdot \tilde{\zeta}^{(2)} \
\zeta^{(2)}\cdot
\tilde{\zeta}^{(3)}$ -- but the interchange of arguments leads to a
change of
sign in the limit that $z_2$ is real.   Thus, the leading behaviour
derived
from summing terms in the full amplitude that include these factors is
suppressed relative to the individual terms.

Cancellations of this type and others that depend on the relative
order of
fermions must be systematically taken into account in adding the
sixteen terms
that make up one cyclic order.
Summing the integrands of all sixteen  terms of the form
(\ref{fourgravscat2})
in the limit  of small $y_2, y_3$ and  $y_4$  gives
\begin{equation}\label{limitfourgrav}
{y_2 y_4 (1+ x_3 x_4) \over (1+ x_4^2)^2 (x_4-x_3)^3 x_3^3 }  \exp(E)+  O
(y_i^3),
\end{equation}
which vanishes quadratically in the $y_i$'s.   This  implies that
the dominant
contribution to the $(1,2,3,4)$ cyclic four graviton amplitude in the
high-energy fixed-angle limit  is proportional to
\begin{equation}\label{s2t2}
  \zeta^{(1)} \cdot \tilde{\zeta}^{(2)} \ \zeta^{(2)}\cdot
\tilde{\zeta}^{(3)}
\  \zeta^{(3)} \cdot \tilde{\zeta}^{(4)}  \ \zeta^{(4)} \cdot
\tilde{\zeta}^{(1)}\  \frac{1}{s^2t^2},
\end{equation}
where the $u^{-1}$ coming from (\ref{stu}) has cancelled the
$k_2\cdot k_4$ in
(\ref{fourgravscat2}) and the extra inverse powers of invariants
are produced
by
the $y_2y_4$ factor in (\ref{limitfourgrav}). The other cyclically
distinct
contractions gives terms proportional
to $s^{-2}u^{-2}$ and $t^{-2}u^{-2}$ respectively.  There are many
other terms
that are suppressed by higher inverse powers of $s$, $t$ or $u$.

There is an apparent divergence in the coefficient of the leading power
behaviour at the  end-point $x_3 = x_4$ in (\ref{limitfourgrav}).
 However,
this is an illusion.  The approximation used in  (\ref{limit2})
assumed that
$|x_i-x_j| \gg y_i$ which is not true in the region that  the
vertices $i$ and
$j$  approach each other  and the boundary simultaneously.     This
is the
boundary of moduli space in which an open-string strip pinches
which may result
in a logarithmic singularity associated with the level-one intermediate
open-string states described earlier.  But
a simple scaling argument shows that  there cannot be any
divergences for the
cyclic
contractions with external gravitons.   This may be seen by
considering  the
limit in which the vertex operators 3 and 4 are  close  to each
other  and  the
boundary using the appropriately rescaled complex  variable,
$\rho_4$, defined
by,
\begin{equation}\label{scaling1}
z_4=x_3+  y_3\rho_4.
\end{equation}
The potential divergence arises as $y_3\to 0$. The measure $d^2z_4\to\int
y_3^2 d^2\rho_4$ so that divergences at the boundary come from
terms which have
$y_3^{-k}$ with $k\leq 3$. It is easy to see that  the prefactor in
(\ref{fourgravscat2})  scales like $\epsilon^{-2}$ and hence there  is no
divergence  in that limit.

When  some or all of  the
external states are \NSNS\ antisymmetric  tensors there are extra
contributions to the amplitude.   The  leading term, of order
$\kappa^2$,  is
now  associated
with a world-sheet that is the product of two disks with a pair of
closed-string states
attached to each.  This contributes  a term proportional to
\begin{equation}\label{twodisk}
\int d^{10}y e^{-2\pi/\kappa} A_{2B}(k_1,k_2;y) A_{2B}(k_3,k_4;y)
e^{i q\cdot
y} ,
\end{equation}
where  $A_{2B}( k_1,k_2,y)$ is the antisymmetric tensor two-point
function and
the $y^\mu$ integral
gives $\delta^{(10)} (\sum_r k_r^\mu + q^\mu)$, where $q^\mu$ is
again the
momentum arising from other particles in the process that soak up
the sixteen
fermionic zero modes.   When the  four antisymmetric tensors are
attached to a
single disk there are  new issues related to the Dirichlet
divergences that
arise with non-cyclic contractions.

\subsection{ Dirichlet divergences.}
Unlike with cyclic contractions, there are contributions to non-cyclic
contractions in which there  are scalar intermediate states
present in some
closed-string channels.   For example, the contribution with kinematic
coefficient $\zeta^{(1)} \cdot \tilde\zeta^{(2)} \  \zeta^{(2)}
\cdot \tilde
\zeta^{(1)} \ \zeta^{(3)} \cdot  \tilde \zeta^{(4)} \  \zeta^{(4)}
\cdot \tilde
\zeta^{(3)}$ involves a set of
of contractions analogous to (\ref{fourgravscat}) but the structure
of the
prefactors is such that there is a logarithmic singularity   from the
integration region at
which $z_3$ and $z_4$  approach each other and the world-sheet boundary
simultaneously (this is made explicit by the rescaling in
(\ref{scaling1})).
 Such divergences, which are generic in  D-instanton processes  are
guaranteed
to cancel \cite{polchinst}.

\medskip
\ifig\ffour{World-sheets of order $\kappa^3$ with Dirichlet
divergences due to
open-string degenerations,  indicated by dashed lines.   (a) The
divergence
arising from the degeneration of the disk which  is cancelled by
(b), a  single
degeneration of the annulus.   The double degeneration of the annulus is
cancelled by other diagrams.   All boundaries are mapped to the
same space-time
point.}
{\epsfbox{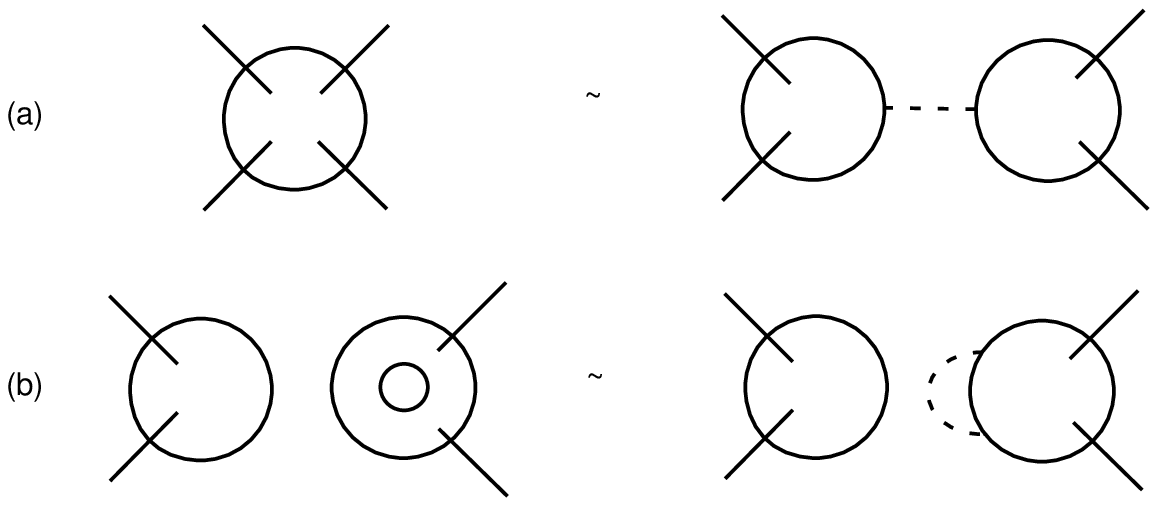}}
\medskip
For example, the divergence  illustrated in \ffour(a) corresponds to a
factorization of the four-point function  on the open string
state  and  has the form
\begin{eqnarray}
&&(k_3+k_4)\cdot (k_1+k_2) \int_a^L {dq\over q}
A_2(k_1,k_2;y)
A_2(k_3,k_4;y)\nonumber \\
&& \quad  = \ln (L/a) {\partial\over \partial
y^\mu}A_2(k_1,k_2;y){\partial\over \partial y_\mu}A_2(k_3,k_4;y),
\label{divlog1}
\end{eqnarray}
where $a$ is an arbitrary constant and $A_2$ is the two-point
function on a
disk for any of the massless closed-string states.

As expected \cite{polchinst,greengas}, the divergence cancels when
account is
taken of  other (disconnected)  diagrams of the same order in
$\kappa$ but with
more boundaries.  In this example the relevant diagram is the
product of an
annulus
and  a
disk   with two states attached to each and all three
boundaries mapped to the same point in the target space, $y^\mu$.
The limit in
which one open-string strip of the annulus degenerates  again leads to a
logarithmic divergence due to the intermediate level-one
open-string vector
state.  This divergent term has the form,
\begin{equation}\label{divlog2}
{1\over 2} \ln (L/b)  \left\{{\partial^2\over \partial
y^2}A_2(k_1,k_2;y)A_2(k_3,k_4;y)+
A_2 (k_1,k_2,y){\partial^2\over
  \partial y^2}A_2(k_3,k_4;y)\right\},
\end{equation}
where $b$ is an arbitrary constant (which need not necessarily be
equal to $a$).
Adding (\ref{divlog1}) and (\ref{divlog2}) give  a total derivative
in $y$ so
the $L$-dependence cancels after integration over the collective
coordinates
and hence the sum of the terms is finite.

\medskip
\ifig\ffive{The disconnected world-sheet in \ffour(b) (a disk and an
 annulus
with coincident boundaries) as seen from the space-time point of view
and has
the structure of a loop diagram correction to  the instanton process. }
{\epsfbox{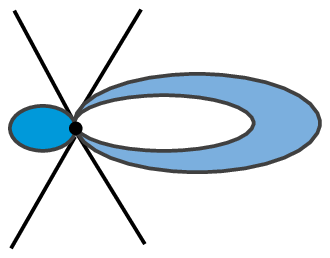}}
\medskip
 It is noteworthy that the annular world-sheet in \ffour(b) is really
a kind of
closed-string loop diagram from the target space point of view since
all three
boundaries coincide. The disconnected world-sheet  of \ffour(b) is
shown from
the  space-time point of view in \ffive, where it looks like a loop
 correction
that is expected in the background of an instanton in field theory.
 Of course,
from the field theoretic point of view such loops are ill-defined in ten
dimensions.  The cancellation of these divergences persists to all orders

The coefficient of $\ln a$ in  (\ref{divlog1}) is of the same form as
(\ref{twodisk}) apart from an overall factor of $\kappa s $.  This
suggests
that in some more systematic treatment of the sum of diagrams these
contributions could come be the first two terms in an expansion of
\begin{equation}
\label{expont}
e^{-2\pi /\kappa} e^{2\pi\kappa s \ln( b/a)},
\end{equation}
which is an exponentially suppressed
contribution to the one-instanton  amplitude.

These Dirichlet divergences arise from a boundary of moduli space
that only
includes part of the region which dominated   in the fixed-angle
limit.  In
other words, cancellation of the divergences does not eliminate the
point-like
behaviour.   This is seen most clearly by considering the
scattering process in
a physical light-cone gauge, as follows.

\vskip0.2cm\noindent
{\it  Point-like scattering in the light-cone gauge}\hfill\break
 The   point-like behaviour of scattering
amplitudes  on a disk is  described in the light-cone gauge following the
analysis in the bosonic string theory in \cite{greenpointlike}.
Recall that
tree-level  elastic scattering of  fundamental
closed-string states  can be described in a physical light-cone
frame  in which
 $\tau =X^+$,
where $X^+$ is time in the infinite momentum frame.   In this
parameterization
the independent variables are, $X^i$  ($i=1,\cdots,8$), the transverse
coordinates.    The scattering process may be mapped to the configuration
   in which the closed-string  world-sheet has width
in $\sigma$ equal to the total $p^+$ carried by the incoming
states,  which  is
conserved throughout the process \cite{mandelstam}.

\medskip
\ifig\fsix{The light-cone gauge parameterization of the
world-sheets describing
the scattering of four closed strings with  (a)   a Neumann
boundary and (b) a
Dirichlet boundary.  Dashed horizontal lines are identified
periodically to form  closed strings.}
{\epsfbox{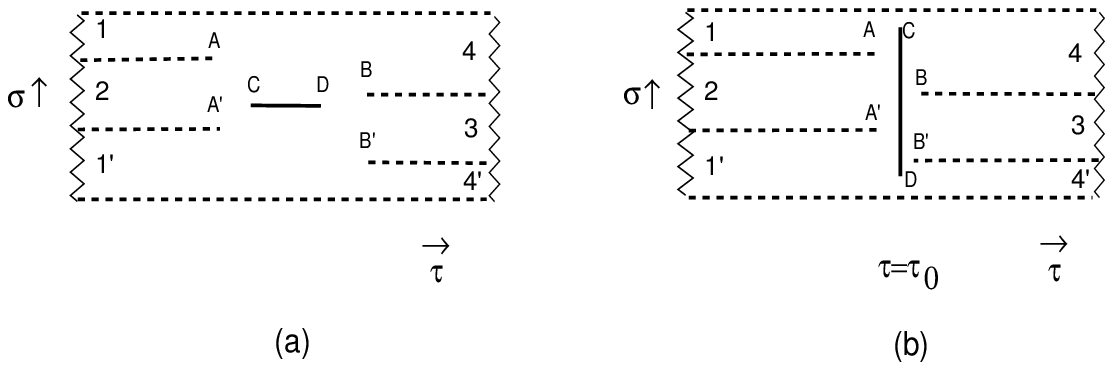}}
\medskip
In the original open-string theory (in which open strings had
Neumann boundary
conditions in all directions) the insertion of a single boundary in the
world-sheet is a unitarity correction (\ffive(a)), which is
represented by the
insertion of
a   boundary at fixed $\sigma$, on which
the coordinates satisfy $\partial_n X^i =0$ (where $\partial_n$
indicates the
normal derivative).   The length of the boundary and its (complex)
position are
the moduli that are to be integrated.  This describes   an
intermediate open string propagating in a closed-string process.

By contrast, the lowest-order contribution to the process in the
presence of a
D-instanton  is represented by the insertion of a boundary
with Dirichlet boundary conditions in all directions.  This is
mapped to a
boundary at {\it fixed} $\tau = y^+$ in the light-cone parameterization
(\ffive(b)).    The boundary condition is now $\partial_t X^i =0$, or
$X^i(\sigma, \tau =y^+) = y^i$.   The length of this boundary and
its (complex)
position are again the  three world-sheet moduli that are to be
integrated
over.  Furthermore, the space-time position of the boundary,
$y^\mu$,  defines
the translational zero modes  of the instanton that form  the
bosonic moduli
that are to be integrated over in performing the sum over all instanton
configurations.    The precise mapping involved   is described in
\cite{greenpointlike}.  It applies equally well to any of the
instantons that
correspond to wrapping the euclidean world-volumes of $p$-branes around
compactified directions.   The difference in those cases is that
$(p+1)$ of the
$X^i$ satisfy Neumann conditions on the boundary and and there are
only $(7-p)$
components to $y^i$.

 This distinction between a unitarity correction and a
D-instanton correction   changes the high energy behaviour of scattering
amplitudes.  The fixed $\tau$ boundary in the case of the
D-instanton represents a  condition imposed on the sum over string
histories that projects onto those histories which have a  point-like
component at transverse position $y^i$ at  the \lq time' $\tau = y^+$ .
Fixed-angle scattering is dominated by those histories in which a finite
fraction of the energy on the incoming strings is concentrated at a
point at
the moment that the strings scatter, leading to power-behaved amplitudes.
The point-like fixed-angle behaviour comes from the integration
region in which
the points $A$ and $B$ in \ffive(b)\ approach the vertical slit.  This is
independent of the  Dirichlet divergence, described earlier,  which
arises
from the end-point at which the vertical slit in \ffive(b)\  spans
the whole
string so that the points $C$ and $D$ touch.

\vskip 0.3cm

The  point-like effects arising from the disk diagram are presumably
 resolved
by accounting for other contributions to the amplitude.  For example,
 in the
calculation of a full amplitude there must be  fermionic moduli
 provided  by
external sources.  Furthermore, as we saw earlier, the presence of
multiply-charged D-instantons is essential for the consistency of the
theory.
We have not considered how fixed-angle scattering might be affected
by such
contributions.   Whereas the  physics of D-particles is controlled by the
eleven-dimensional Planck scale \cite{pouliot,ferretti,douglas},
 the only scale that
enters in the  multi D-instanton moduli space is the
ten-dimensional Planck
scale \cite{gutgreenus} which is larger at weak coupling.  It would be
interesting to understand how this apparent distinction in scales
is reflected
in the physical properties of the theory.

\vskip 0.3cm

{\bf Note}:  While this paper was being prepared several papers
\cite{periwal,kawai,tseytlina,barbonb} appeared in which related
interesting
issues  concerning  D-instantons are discussed.

\vskip 0.8cm
\noindent{ \Large{\bf Acknowledgments}}:

\noindent   MBG is  grateful to   the Ecole Normale for hospitality
in August
1996 where parts of this work were carried out.

 \vskip 0.8cm
\noindent{ \Large{\bf APPENDIX}: {\bf The SUGRA instanton action}}
\vskip 0.3cm
The strategy  is to transform the \RR scalar, $C^{(0)}$,  to an
eight-form
potential, $C^{(8)}$, with field strength $F_9$.    There are
several arguments
for using such a formulation of the IIB theory.\footnote{Much of
this appendix
is based on discussions with Gary Gibbons}   Firstly, the
continuation from
minkowskian to euclidian signature is very simple since the Maxwell-type
lagrangian density, $F_9^2$, does not transform.   Furthermore, the
euclidean
action is manifestly positive in this dual formulation, without the
introduction of  a boundary term.    We will see that this implies  the
presence of a boundary term in the scalar field part of the usual
type IIB
euclidean action, which makes it positive.

In order to motivate the form of the dual action we start by
considering the
first-order minkowskian action  (it will be sufficient to consider
the scalar
field part of the action  with a  flat metric
since the
metric is trivial in the instanton solution  in the Einstein frame
\cite{greenc}),
\begin{equation}
\label{dualact}
S(P^{(9)},C^{(0)}) =  \quart \int   \left(d\phi \wedge *d\phi +
e^{-2\phi}
P^{(9)}\wedge *P^{(9)}   - 2   C^{(0)} \wedge dP^{(9)} \right),
\end{equation}
where $P^{(9)}$ is a nine-form field.  The functional integral is
of the form
$\int DP^{(9)} DC^{(0)} e^{iS}$ and we will write $S = S_M$   when the
signature is minkowskian and $S= i S_E$  when it is euclidean.

  Later we will see that performing the integration over $P^{(9)}$
first gives
rise to the $C^{(0)}$ kinetic terms in the IIB action together
with a specific
surface term.   But to begin with we will consider  integrating
over $C^{(0)}$
first, which   imposes the constraint $dP^{(9)} =0$ that  is solved
by $P^{(9)}
= d C^{(8)}$ (ignoring global issues).   Substituting this in the action
(\ref{dualact}) gives,
\begin{equation}
\label{ceightact}
S(C^{(8)}) = \quart \int   \left(d\phi \wedge *d\phi + e^{-2\phi} d
C^{(8)}
\wedge * d C^{(8)}   \right).
\end{equation}
With Minkowski signature this can be expressed as
\begin{equation}
\label{eightmink}
S_{M}(C^{(8)}) = \quart   \int  \left(d\phi  -  e^{-\phi} *d
C^{(8)}  \right)
\wedge * \left(d\phi  + e^{-\phi} *d C^{(8)}  \right)
\end{equation}
where we have made use of the   properties of a $p$-form,
$a^{(p)}$, under
Poincar\'e duality in ten dimensions,
\begin{equation}
\label{minkdual}
**a^{(p)} = (-1)^{p+1} a^{(p)}.
\end{equation}

With Euclidean signature Poincar\'e duality implies,
\begin{equation}
\label{eucldual}
* *a^{(p)} = (-1)^p a^{(p)},
\end{equation}
and the action (\ref{ceightact}) becomes
\begin{equation}
S_{E} (C^{(8)}) =  \quart \int   \left(d\phi \pm e^{-\phi} * d
C^{(8)} \right)
\wedge *\left(d\phi \pm e^{-\phi} * d C^{(8)} \right)
   \pm\half  \int e^{-\phi} d\phi \wedge d C^{(8)}.
\label{eucdual}
\end{equation}
The first term on the right-hand side is positive semi-definite while the
second term is a total derivative that can be written as,
\begin{equation}
\label{secterm}
\int e^{-\phi} d\phi \wedge d C^{(8)} = - \int d\left(e^{-\phi} d C^{(8)}
\right) = - \oint_{\partial M} e^{-\phi} F_9,
\end{equation}
where $F_9 = d C^{(8)}$ and   $\partial M= \partial{M_\infty}+
\partial{M_0}$
denotes the boundary of space-time which  consists of the  $S^9$ at
$r=\infty$
($\partial {M_\infty}$) and the $S^9$ around $r=0$
($\partial {M_0}$).  From
this it follows that the euclidean action satisfies a Bogomol'nyi bound,
\begin{equation}
\label{bogom}
S_{E} (C^{(8)}) \ge \half \left| \oint_{\partial M} e^{-\phi}
F_9\right |.
\end{equation}
The action is minimized and the  bound saturated when the first term in
(\ref{eucdual}) vanishes, so that
\begin{equation}
\label{bogeqs}
d\phi  \pm e^{-\phi} * d C^{(8)} =0,
\end{equation}
are satisfied by the BPS and anti-BPS solutions.  This is the dual
form of
(\ref{susyleft}).
The   \lq electric' charge on the instanton is given by the topological
integral $\oint_{\partial M_\infty} F_9/2 = 2\pi  q$, whereas in
the original
formulation it is equal to the integral of the Noether charge for
the $C^{(0)}$
shift symmetry.     This gives the instanton action,
\begin{equation}
\label{instactsol}
S^{(q)} = {2\pi \over \kappa}  |q|.
\end{equation}
The parameter $q$ becomes quantized in integer units  by the
Dirac--Nepomechie--Teitelboim
argument for a instanton in the presence of a euclidean seven-brane.

In order to see how the euclidean IIB action arises from
(\ref{dualact})  we
will now consider performing the $P^{(9)}$ integration first.  This
is obtained
by writing the action as the quadratic form,
\begin{eqnarray}
S_E(P^{(9)},C^{(0)}) & =&  \quart  \int   \left(   d\phi \wedge * d\phi +
   e^{-2\phi} \left( P^{(9)} +  i e^{2\phi} *dC^{(0)} \right) \wedge *
\left(P^{(9)} + i e^{2\phi}  * dC^{(0)}\right)  \right. \nonumber \\
&& \left. +   e^{2\phi} dC^{(0)} \wedge * d C^{(0)} -  2 d\left(i C^{(0)}
 P^{(9)} \right)  \right),
\label{czeroact}
\end{eqnarray}
where the factors of $i$ arise from the Wick rotation of the first
two terms on
the right-hand side of (\ref{dualact}) together with the fact that
$S_E = - i
S$.
Shifting variables  to $P^{'(9)} = ( P^{(9)} + i e^{2\phi}
*dC^{(0)})$ and
performing the  $P^{'(9)}$ integration in the functional integral
gives the
scalar field part of the IIB supergravity action together with a
surface term,
$ \oint_{\partial M} d( e^{2\phi} C^{(0)} \wedge * dC^{(0)})$.
This  shift of
variables is only possible if $dC^{(0)} = i df$ is imaginary, as
anticipated
earlier,  in which case   the action is given by,
\begin{equation}
\label{factt}
S_{E}(f)  =  \quart \int  \left(  \left(d\phi \wedge * d\phi - e^{2\phi}
df\wedge * df  \right)  - 2 d\left(e^{2\phi} f \wedge * df\right)\right),
\end{equation}
which  gives rise to equations   of motion  (\ref{susyleft}) with
instanton
solutions  (\ref{sol}) \cite{greenc} in which  $f = A + e^{-\phi}$,
where $A$
is an arbitrary constant.   The action for a charge $q$ instanton,
obtained by
substituting the solution into (\ref{factt}), comes entirely from
the surface
term which gets contributions from  $\partial {M_\infty}$  and
$\partial {M_0}$.   Using the explicit solution for $e^\phi$ the
boundary term
at $r=0$ can be shown to equal $2\pi(A-1/\kappa)q$ while that at
$r=\infty$ is
equal to $2\pi Aq$ so that the action is again given by
(\ref{instactsol}).  If
we choose $A=0$  the action comes entirely from the surface at
$r=\infty$.

The presence of the boundary term in the scalar action affects the
functional
integral in that it restricts the class of functions, $f$,  to
those that have
continuous derivatives (in other words, the action is only additive
for this
class of functions).  From the duality relation $F_9 \sim e^\phi
df$ this is
consistent with the fact that  $F_9$ is continuous in the absence of
seven-brane sources.


\begin{thebibliography}{99}
\bibitem{stromingerbecker} K. Becker, M. Becker and A. Strominger, {\it
Fivebranes, membranes and nonperturbative string theory},
hep-th/9507158,
  Nucl. Phys. {\bf B456} (1995) 130.
\bibitem{ooguri} H. Ooguri and C. Vafa, \it Summing up D instantons,
  \rm HUTP-96-A036, { hep-th/9608079}.
\bibitem{harveymoore} J.A. Harvey and G. Moore, \it Five-brane
  instantons and $R^2$ couplings in N=4 string theory, \rm EFI-96-38,
  { hep-th/9610237}.
\bibitem{greengutperlec} M.B. Green and M. Gutperle, \it Light cone
  supersymmetry and D-branes, \rm  hep-th/9604091, Nucl. Phys. {\bf
  B476} (1996) 484.
 \bibitem{greenpointlike}  M.B. Green, \it Point-like structure and
  off-shell dual amplitudes, \rm  Nucl. Phys. {\bf B124} (1977) 461.
\bibitem{greendual} M.B.  Green, {\it Space-time duality and
Dirichlet string
theory},  Phys.  Lett. {\bf B266} (1991) 325.
\bibitem{gutperlea} M. Gutperle, \it Multiboundary effects in
  Dirichlet string theory, \rm hep-th/9502106, Nucl. Phys. {\bf B444}
  (1995), 487.
\bibitem{veneziano} G. Veneziano, \it Construction of a crossing
  symmetric, Regge behaved amplitude for linearly rising
  trajectories, \rm  Nuov. Cim. {\bf 57a} (1968) 190.
\bibitem{amati} V. Allesandrini, D. Amati and B. Morel, \it The
  asymptotic behavior of the dual pomeron amplitude, \rm  Nuov.
Cim. {\bf 7A}
(1971) 797.
\bibitem{grossmende} D.J. Gross and P.F. Mende, \it String theory beyond
  the Planck scale, \rm Nucl. Phys. {\bf B303} (1988)
407.
\bibitem{greenc} G.W. Gibbons, M.B. Green and M.J. Perry, \it
  Instantons and seven-branes in type IIB superstring theory, \rm
hep-th/9511080,   Phys. Lett. {\bf B370} (1996) 37.
\bibitem{wittenboundstates} E. Witten, \it Bound states of strings and
  p-branes, \rm hep-th/9510135,   Nucl. Phys. {\bf B460} (1996) 335.
\bibitem{pouliot}    D.  Kabat and P. Pouliot, {\it A comment on
zero-brane
quantum mechanics}, hep-th/9603127, Phys.  Rev.  Lett. {\bf 77}
(1996) 1004.
\bibitem{ferretti}    U.H. Danielsson, G. Ferretti,   B. Sundborg,
{\it D
particle dynamics and bound states}, hep-th/9603081, Int. J. Mod.
Phys. {\bf
A11} (1996) 5463.
\bibitem{douglas} M.R. Douglas, D. Kabat, P. Pouliot and
  S.H. Shenker, \it D-branes and short distances in string theory,
  \rm RU-96-62, hep-th/9608024.
\bibitem{thooft} G. t Hooft, \it Computation of the quantum effects
  due to a four-dimensional pseudoparticle, \rm  Phys. Rev. {\bf D14}
  (1976) 3432.
\bibitem{grisaru}M.T.~Grisaru , A.E.M~Van de Ven and D.~Zanon,  {\it
Two-dimensional
    supersymmetric sigma models on Ricci flat Kahler manifolds are  
not finite},
   Nucl.  Phys.  {\bf B277} (1986) 388 ; {\it Four loop divergences  
for the N=1
supersymmetric
    nonlinear sigma model in two-dimensions}; Nucl.  phys. {\bf  
B277} (1986)
409.
\bibitem{wittengross} D.J. Gross and E. Witten, \it Superstring
  modifications of Einstein's equations \rm Nucl. Phys. {\bf B277}
  (1986) 1.
\bibitem{greenschwarziii} M.B.  Green and J.H.  Schwarz, {\it
Supersymmetric
dual string theory III}, Nucl.  Phys. {\bf B198} (1982) 441.
\bibitem{polchinst} J. Polchinski, {\it Combinatorics of  boundaries
in string
theory},   hep-th/9407031, Phys.  Rev. {\bf D50} (1994) 6041.
\bibitem{greengas} M.B. Green, {\it A gas of D-instantons}, Phys.
Lett. {\bf
B354}
(1995); {\it Boundary effects in string theory}, contributed to
STRINGS 95:
Future Perspectives in String Theory, USC, March 1995, hep-th/9507121.
\bibitem{schwarz} J.H. Schwarz, \it Covariant field equations of
  chiral N=2 D=10 supergravity\rm, Nucl. Phys. {\bf B226} (1983) 269.
\bibitem{greeniib} M.B. Green, {\it Point-like states for type IIB
superstrings},   hep-th/9403040,
Phys.  Lett. {\bf B329}  (1994) 435.
\bibitem{nilsson} M. Cederwall, A. von Gussich, B.E.W. Nilsson and
  A. Westerberg, \it The Dirichlet super three brane in
  ten-dimensional type IIB supergravity, \rm GOTEBORG-ITP-96-13,
  hep-th/9610148; M. Cederwall et al., \it The Dirichlet super p-branes
  in ten dimensional type IIA and type IIB supergravity, \rm
  GOTEBORG-ITP-96-14, hep-th/9611159.
\bibitem{schwarzb} M. Aganagic, C. Popescu and  J.H. Schwarz, \it
  D-brane actions with local kappa symmetry\rm ,  CALT-68-2081A,
  hep-th/9610249; \it Gauge invariant and gauge fixed D-brane
  actions, \rm CALT-68-2088, hep-th/9612080
\bibitem{townsend} E. Bergshoeff and P.K. Townsend, \it Super
  D-branes, \rm DAMTP-R-96-53, hep-th/9611173.
\bibitem{rubakov} V.A. Rubakov and M.E. Shaposhnikov, \it  Electroweak
  baryon number nonconservation in the early universe and in
  High-energy collisions, \rm CERN-TH-96-13,  hep-ph/9603208.
\bibitem{tseytlinb} A.A. Tseytlin, {\it Heterotic Type I superstring
  duality and low energy effective action},  hep-th/9512081,
  Nucl. Phys. {\bf B467} (1996) 383.
\bibitem{gutgreenus} M.B.  Green and M. Gutperle, {\it Configurations
of two
D-instantons},  DAMTP-96-110, hep-th/9612127.
\bibitem{zagiera} D. Zagier, {\it Eisenstein series and the Riemann
    zeta function } in Tata Inst. Fund. Res.
Studies in Math., 10, Tata Inst. Fundamental Res., Bombay, 1981.
\bibitem{tarrasa} A. Terras, {\it Harmonic analysis on symmetric
    spaces and applications I}, Springer, New York-Berlin, 1985.
\bibitem{klebanova} I.R. Klebanov and L. Thorlacius, \it The size of
  p-branes, \rm hep-th/9510200, Phys. Lett. {\bf B371} (1996) 51.
\bibitem{gutthesis} M. Gutperle, Ph.D thesis, University of Cambridge,
  (unpublished).
\bibitem{barbon} J.L.F. Barbon,  {\it D-brane form-factors at high
  energy}, hep-th/9601098, Phys. Lett. {\bf B382} (1996) 60.
\bibitem{mandelstam} S. Mandelstam, \it Interacting string picture of
  dual resonance models, \rm Nucl. Phys. {\bf 64} (1973) 205.
\bibitem{periwal}
V. Periwal,  \it Matrices on a point as the theory of everything,
\rm PUPT-1665, hep-th/9611103; \it Antibranes and crossing symmetry, \rm
PUPT-1676,
hep-th/9612215.
\bibitem{kawai}  N. Ishibashi, H. Kawai,Y. Kitazawa and A. Tsuchiya,
  \it A large N reduced Model as superstring, \rm KEK-TH-503,
hep-th/9612115.
\bibitem{tseytlina} A.A. Tseytlin, \it Type IIB instanton as a wave in
  twelve dimensions, \rm  CERN-TH-96-333, hep-th/9612164.
\bibitem{barbonb}
J.L.F. Barbon, {\it Fermion exchange between D-instantons},
CERN-TH-96-360, hep-th/9701075.
\end{thebibliography}
\end{document}